\documentclass[12pt]{article}
\usepackage[margin=0.8in]{geometry}
\usepackage{amsmath,amssymb,amsthm,bm,mathtools}
\usepackage{natbib}
\usepackage{booktabs}
\usepackage{enumitem}

\usepackage{dsfont}

\usepackage{color,soul}
\soulregister\cite7
\soulregister\citep7
\soulregister\citet7
\soulregister\ref7

\usepackage{outlines}
\usepackage[justification=centering]{caption}
\usepackage{subcaption} 
\usepackage{hyperref}
\hypersetup{colorlinks=true,citecolor=blue,linkcolor=blue,urlcolor=blue}
\usepackage{bm}
\usepackage{bbm}
\usepackage{textcomp}
\usepackage{xcolor}
\usepackage[nocomma]{optidef}
\usepackage{tabularx}
\usepackage{comment}
\usepackage{tikz}
\usepackage{algorithm,algpseudocode}    

\newcommand{\bs}[1]{\boldsymbol{#1}}

\newcommand{\ignore}[1]{}
\theoremstyle{plain}
\newtheorem{theorem}{Theorem}[section]
\newtheorem{lemma}[theorem]{Lemma}

\newtheorem{proposition}[theorem]{Proposition}

\theoremstyle{definition}

\theoremstyle{remark}
\newtheorem{remark}[theorem]{Remark}

\title{Global Testing of Gaussian Means by Combining Marginal and Conditional Evidence}
\author{
  Deepra Ghosh\thanks{Department of Statistics, Operations and Data Science,
  Temple University.} \qquad
  Edgar Dobriban\thanks{Department of Statistics and Data Science,
  University of Pennsylvania.} \qquad
  Sanat K. Sarkar\footnotemark[1]
}
\date{October 2026}

\begin{document}
\maketitle
\begin{abstract}
We propose a test of the global null hypothesis of zero means in correlated multivariate Gaussian data by combining marginal and conditional evidence at each coordinate. The conditional tests are based on the distribution of each coordinate given the remaining coordinates. At each coordinate, the corresponding two-sided $p$-values are combined using the Simes method, followed by a Bonferroni correction across coordinates. Because conditioning can either strengthen or weaken evidence against the null, neither marginal nor conditional testing is uniformly preferable. The proposed procedure adapts to their relative strengths while retaining the simplicity of minimum-type global testing. We establish finite-sample control of the global Type-I error under both known covariance and covariance estimated from an independent sample. The latter setting requires a careful analysis of the dependence between studentized marginal and conditional statistics. We evaluate the power of the proposed test through simulations, compare its performance with existing global tests, and illustrate its practical utility using a TCGA carcinoma dataset. 
\end{abstract}


\section{Introduction}\label{sec:introduction}


Testing the mean of a Gaussian vector is a fundamental problem in multivariate statistics, with applications spanning biology, economics, finance, and other fields \citep[see, e.g.,][]{ChenQin2010,GibbonsEtAl1989,FanLiaoYao2015}. A wide range of testing procedures has accordingly been developed to address this problem under different statistical settings \citep[see, e.g.,][]{Hotelling1931,DonohoJin2004,Zhang2017,FrostigBenjamini2022,FengJiangLiLiu2024}. 


In this work, we consider testing whether the mean of a Gaussian vector is zero against the alternative that it is nonzero. Our approach exploits two complementary sources of evidence: the marginal mean of each coordinate and its conditional mean after linearly adjusting for the remaining coordinates. These capture distinct aspects of the underlying signal, as a coordinate may exhibit a strong marginal effect but a weak conditional effect, or vice versa. This distinction motivates the development of global testing procedures that combine marginal and conditional evidence to improve the detection of departures from the null. 
 


Specifically, we consider the Gaussian model
\begin{equation} \label{eq:gaussian-model}
\bs X\sim N_d(\bs\mu,\bs\Sigma),
\end{equation}
where \(\bs\Sigma\) is positive definite, and test the global null hypothesis $H_0:\bs\mu=\bs0$ against $H_1:\bs\mu\neq\bs0$. The two sources of evidence in our approach correspond to the coordinates of the mean vector $\bs\mu$ and its precision-adjusted counterpart $\bs\delta=\bs\Sigma^{-1}\bs\mu$, respectively. 

A natural starting point for global testing is the Bonferroni procedure, which rejects the null hypothesis when the smallest marginal $p$-value does not exceed $\alpha/d$ \citep{Bonferroni1936}. 
Classical global testing procedures aggregate evidence in different ways. Fisher's combination test sums transformed $p$-values \citep{Fisher1932}, Tippett's procedure focuses on the smallest $p$-value \citep{Tippett1931}, and Stouffer's method combines inverse-normal transformations of $p$-values \citep{StoufferEtAL1949}. An alternative approach uses quadratic forms of the observations, as in the Mahalanobis test under known covariance \citep{Mahalanobis1936} and Hotelling's $T^2$ test with an independently estimated Wishart covariance matrix \citep{Hotelling1931}.


The choice of aggregation rule is especially consequential when departures are sparse. Building on the Gaussian detection framework of \citet{DonohoJin2004}, \citet{AriasCastroEtAl2011} studied ANOVA, maximum-type tests, and higher criticism under sparse regression alternatives. Their results illustrate the distinct detection regimes of quadratic, maximum, and higher criticism procedures. In particular, maximum-type tests can be asymptotically optimal in very sparse regimes, whereas higher criticism can improve detection at intermediate sparsity levels. 

Dependence creates additional challenges.
The Simes test \citep{Simes1986} rejects when
\(
\min_{1\le k\le d}\) \(\frac{dP_{(k)}}{k}\le\alpha,
\)
where \(P_{(1)}\le\cdots\le P_{(d)}\). It has exact size \(\alpha\) for independent continuous uniform \(p\)-values, and its validity under various forms of positive dependence has been studied in \citet{HochbergRom1995,SamuelCahn1996,SarkarChang1997,Sarkar1998,BenjaminiYekutieli2001,Sarkar2002,Sarkar2008a,Sarkar2008b}. 
Separately, \citet{Sidak1967} established a valid minimum-\(p\) cutoff for two-sided Gaussian testing without restrictions on correlation signs. 

Several recent developments seek greater power by exploiting covariance information, signal sparsity, or both. \citet{Zhang2017} constructed precision-adjusted mean tests, including the maximum-type Z-Max procedure and the adaptive Z-TopK procedure, which aggregates leading adjusted coordinates. 
These methods exploit the possibility that covariance adjustment can amplify sparse signals. 
\citet{HallJin2010} similarly argued for the benefits of transforming correlated observations for sparse-signal detection. 
Other approaches include \citet{FengJiangLiLiu2024}, 
who combined sum-of-squares and maximum statistics, 
and
\citet{QiuChenQiu2026}, who combined quadratic, maximum, and higher criticism components. 
\citet{ZhangWangShao2025} constructed adaptive tests based on combinations of even-order U-statistics.


Covariance-informed global testing has been investigated by \citet{FrostigBenjamini2022},
who proposed a Simes–Hotelling procedure that applies Hotelling tests to randomly sampled coordinate subsets and combines their \(p\)-values. 
The Cauchy combination test of \citet{LiuXie2020} provides another approach with asymptotic validity. 


In this work, we also exploit covariance information, but through a different approach. We propose an evidence-adaptive (EA) global test that first combines the marginal and conditional $p$-values at each coordinate using Simes' method and then applies a Bonferroni correction across coordinates. Our main theoretical contribution is to establish Type-I error control for this procedure under Gaussian models with known covariance and, separately, with covariance estimated from an independent sample. The latter setting poses additional challenges because the marginal and conditional test statistics involve random denominators. We address these challenges by establishing a conditional Simes inequality that underpins the validity of the proposed test.


We empirically investigate the power of the proposed EA test against a broad range of existing methods under various mean and covariance structures. Our numerical results show that EA is generally competitive and can offer substantial power gains in certain settings. In particular, it performs well when two coordinates have marginal and conditional effects of comparable magnitudes, with matching signs for one coordinate and opposing signs for the other. Further improvements are observed under heterogeneous signals within correlated modules and localized peaks with moving-average noise, as examined in Sections~\ref{subsec:correlated-module} and~\ref{subsec:moving average}, respectively. We also illustrate the proposed method using a TCGA carcinoma dataset.

The remainder of the paper is organized as follows. Section \ref{sec:methods} introduces the testing framework and relevant properties of the Simes procedure. Sections \ref{sec:known-covariance} and \ref{sec:unknown covariance} develop the proposed EA procedure and establish its finite-sample validity under known and independently estimated covariance matrices, respectively. Section \ref{sec:simulations} presents simulation studies comparing the proposed procedure with existing global tests, while Section \ref{sec:real data analysis} illustrates its application to TCGA carcinoma data. Section \ref{sec:discussion} concludes with a discussion of the main findings and potential extensions. Proofs of the results for unknown covariance are provided in Appendix \ref{app:proofs}, with supporting technical lemmas and additional simulation results presented in the Supplementary Material (Section \ref{sec:supp}). 


\section{Methodology}\label{sec:methods}
Here, we describe the two settings considered in this work, corresponding to known and unknown covariance matrices.


\subsection{Known and Unknown Covariance Settings}\label{subsec:sampling-framework}

\paragraph{Setting K: Known Covariance Matrix.}
In our first setting, we consider
model~\eqref{eq:gaussian-model}, where $\bs\Sigma$ is a known positive definite covariance matrix. 

\paragraph{Setting W: Unknown covariance matrix.}
In our second setting, 
we consider
model~\eqref{eq:gaussian-model}, where $\bs\Sigma$ is not known. 
However, in addition to $\bs X$ in~\eqref{eq:gaussian-model}, we observe an independent Gaussian sample of size $n+1$ with covariance $\bs\Sigma$. Its sample scatter matrix $\bs A$ satisfies
\begin{equation}\label{eq:wishart-model}
\bs A\sim W_d(n,\bs\Sigma),
\qquad \bs X\perp\bs A.
\end{equation}
Since $E(\bs A)=n\bs\Sigma$, the matrix $\bs A/n$ provides an unbiased estimator of $\bs\Sigma$.



\subsection{Marginal and Precision-adjusted Signal Representations}
\label{subsec:complementary-evidence}

Consider the precision matrix 
$\bs\Omega=(\omega_{ij})_{i,j=1}^d=\bs\Sigma^{-1}$ and the precision-adjusted mean $\bs\delta=\bs\Omega\bs\mu$.
For each $i \in \{1, \ldots, d\}$, we partition the data vector and covariance matrix, respectively,
as \begin{equation}\label{eq:partitioned-matrix-and-vector-1} \bs X = \begin{pmatrix}
X_i \\ \bs X_{-i} \end{pmatrix}, \qquad \bs \Sigma = \begin{pmatrix} \Sigma_{ii} & \bs \Sigma_{i,-i}\\ \bs \Sigma_{-i,i}& \bs \Sigma_{-i,-i} \end{pmatrix}, \end{equation} 
and define $\bs\beta_i =\bs\Sigma_{-i,-i}^{-1}\bs\Sigma_{-i,i}$, the vector of regression coefficients obtained by linearly projecting $X_i$ onto $\bs X_{-i}$, and $v_i=\Sigma_{ii}-\bs\Sigma_{i,-i} \bs\Sigma_{-i,-i}^{-1}\bs\Sigma_{-i,i} =\frac1{\omega_{ii}}$, the corresponding residual variance. Under (\ref{eq:gaussian-model}), we have 
\begin{equation}\label{eq:precision-residual}
X_i=\bs\beta_i^\top\bs X_{-i}
+v_i\delta_i+\varepsilon_i,
\qquad
\varepsilon_i\sim N(0,v_i),
\qquad \varepsilon_i\perp\bs X_{-i}.
\end{equation}
While $\mu_i = E(X_i)$ is the marginal mean, $v_i\delta_i$ is the mean remaining after the linear adjustment using all other coordinates.

\subsection{Simes Combination}
\label{subsec:adaptive-recalibration}

In either setting (K or W), let $P_i$ denote a marginal $p$-value
for $\mu_i=0$ and $Q_i$ its precision-adjusted counterpart for
$\delta_i=0$. Their specific definitions are given in
Sections~\ref{sec:known-covariance} and~\ref{sec:unknown covariance}.

We combine these two $p$-values using the combination rule introduced by
\citet[Section~2, p.~751]{Simes1986}, which, for $m$ ordered $p$-values
$p_{(1)}\le\cdots\le p_{(m)}$, is defined as
$\min_{1\le k\le m}\{mp_{(k)}/k\}$.
Applying this rule to the pair $(P_i,Q_i)$ gives
\begin{equation}\label{eq:combined-p-value}
S_i=\min\{2\min(P_i,Q_i),\max(P_i,Q_i)\},
\qquad i=1,\ldots,d.
\end{equation}
Equivalently,
$$
S_i=
\begin{cases}
2Q_i, & Q_i<P_i/2,\\
P_i, & P_i/2\le Q_i\le P_i,\\
Q_i, & P_i<Q_i<2P_i,\\
2P_i, & Q_i\ge2P_i.
\end{cases}
$$
Thus, sufficiently strong precision-adjusted evidence
($Q_i<P_i/2$) reduces the marginal $p$-value from $P_i$ to $2Q_i$.
When $P_i/2\le Q_i\le P_i$, the combination leaves $P_i$ unchanged.
Weaker precision-adjusted evidence ($Q_i>P_i$) increases the
combined value relative to $P_i$, with a penalty bounded by a
factor of two. In particular, $S_i\le\min\{1,2P_i\}$.


\subsection{A Common Global Test and Its Calibration Requirement}
\label{subsec:common-global-test}

Applying Bonferroni across the combined values gives
\begin{equation}\label{eq:ea-global-p-value}
P_{\mathrm{EA}}=\min\left\{1,\ d\min_{1\le i\le d}S_i\right\}.
\end{equation}
The level-$\alpha$ procedure rejects $H_0$ when $P_{\mathrm{EA}}\le\alpha$. Equivalently, rejection occurs whenever, for some $i$,
$P_i\le\frac\alpha{2d}$, or $Q_i\le\frac\alpha{2d}$, or $\max(P_i,Q_i)\le\frac\alpha d$.
The procedure accepts either sufficiently strong evidence from one representation or moderately strong evidence from both at the same coordinate.

If $S_i$ are valid p-values under the global null, then 
the Bonferroni inequality 
implies that $P_{\mathrm{EA}}$ is also valid.
The same argument holds
conditionally whenever $S_i$ are valid $p$-values under a common conditioning variable.


\section{Known Covariance Structure}\label{sec:known-covariance}

We now develop Setting K in detail as a warmup.

\subsection{Exact Gaussian Marginal and Precision-adjusted Evidence}
\label{subsec:known-inputs}

Define
$Z_{\mathrm M,i}=\frac{X_i}{\sqrt{\Sigma_{ii}}}$
and
$Z_{\mathrm{PA},i}=\frac{(\bs\Omega\bs X)_i}{\sqrt{\omega_{ii}}}$.
We can also write
$Z_{\mathrm{PA},i}=\frac{X_i-\bs\beta_i^\top\bs X_{-i}}{\sqrt{v_i}}$.

Since $\bs\Omega\bs X\sim N_d(\bs\delta,\bs\Omega)$, 
we can  define 
the two-sided 
$p$-values
\begin{equation}\label{eq:known-paired-p-values}
P_i=\bar\Psi_1(Z_{\mathrm M,i}^2),
\qquad
Q_i=\bar\Psi_1(Z_{\mathrm{PA},i}^2),
\end{equation}
where $\bar\Psi_1=1-\Psi_1$ is the survival function of $\chi_1^2$ distribution. The first p-value is uniform when $\mu_i=0$, the second  when $\delta_i=0$.
Inserting~\eqref{eq:known-paired-p-values} into~\eqref{eq:combined-p-value} and then~\eqref{eq:ea-global-p-value} specifies the known-covariance test completely.

\subsection{Pairwise Gaussian Dependence and Finite-Sample Validity}
\label{subsec:known-validity}

Under the global null,
$$
\begin{pmatrix}Z_{\mathrm M,i}\\ Z_{\mathrm{PA},i}\end{pmatrix}
\sim N_2\!\left[
\begin{pmatrix}0\\0\end{pmatrix},
\begin{pmatrix}1&\rho_i\\\rho_i&1\end{pmatrix}
\right],
\qquad
\rho_i=\frac1{\sqrt{\Sigma_{ii}\omega_{ii}}}\in(0,1].
$$
The covariance identity
$\textrm{Cov}
(\bs X,\bs\Omega\bs X)=\bs I$
gives the within-pair correlation.
Therefore, 
the required superuniformity of $S_i$ follows from \citet{SamuelCahn1996} and \citet{SarkarChang1997,Sarkar1998}.


\section{Unknown Covariance}
\label{sec:unknown covariance}

We next develop Setting W, the case of an unknown covariance matrix, under~\eqref{eq:gaussian-model} and~\eqref{eq:wishart-model}, with $n\ge d$ and $\nu=n-d+1$. 

\subsection{Exact Marginal and Precision-adjusted Student-\texorpdfstring{$t$}{t} Evidence}
\label{subsec:studentized-evidence}
We consider the the marginal statistic 
$T_{\mathrm M,i}=\frac{\sqrt n\,X_i}{\sqrt{A_{ii}}}$, 
where we standardize using the independent estimate $A_{ii}/n$ of the variance. 
To estimate the precision adjustment, we partition \(\bs X\) and \(\bs A\) similar to \eqref{eq:partitioned-matrix-and-vector-1} and define
$\widehat{\bs\beta}_i=\bs A_{-i,-i}^{-1}\bs A_{-i,i}$,
$D_i=A_{ii}(1-\widehat{R}_i^2)=\frac{1}{(\bs{A}^{-1})_{ii}}$, and
$h_i=\bs X_{-i}^\top\bs A_{-i,-i}^{-1}\bs X_{-i}$,
where 
\(\widehat{R}_i \equiv \widehat{R}_i(\bs A) = \sqrt{\frac{\bs A_{-i,i}^\top \bs A_{-i,-i}^{-1}\bs A_{-i,i}}{A_{ii}}} \) is the estimated  multiple correlation between $X_i$ and $\bs{X}_{-i}$ based on $\bs{A}$. 
Then, we consider the precision-adjusted statistic 
$T_{\mathrm{PA},i}=\frac{X_i-\widehat{\bs\beta}_i^\top\bs X_{-i}}{\sqrt{(D_i/\nu)(1+h_i)}}$. 
Clearly the marginal statistic has a $t$-distribution, and we will show that the same is true for the precision-adjusted one as well.
Therefore, we will use 
the paired $p$-values
\begin{equation}\label{eq:paired-p-values}
P_i=\bar{\Psi}_{1,n}(T_{\mathrm M,i}^2),
\qquad
Q_i=\bar{\Psi}_{1,\nu}(T_{\mathrm{PA},i}^2),
\end{equation}
where $\bar{\Psi}_{1,k} = 1 - \Psi_{1,k}$ is the survival function of $F_{1,k}$, the $F$ distribution with $1$ and $k$ degrees of freedom. 
When $d=1$, we have $D_1=A_{11}$, $h_1=0$, and $\nu=n$; hence $P_1=Q_1$.

The following basic result records the validity of the two $p$-values.

\begin{proposition}[Validity of the two p-values]\label{prop:exact-inputs}
Under the Gaussian--Wishart model specified by~\eqref{eq:gaussian-model} and~\eqref{eq:wishart-model}, $T_{\mathrm M,i}^2\sim F_{1,n}$ whenever $\mu_i=0$. Whenever $\delta_i=0$,
\begin{equation}\label{eq:conditional-pa-t}
T_{\mathrm{PA},i}^2\mid
(\bs X_{-i},\bs A_{-i,-i})\sim F_{1,\nu}.
\end{equation}
Thus, $P_i$ is exactly uniform under its marginal null, and $Q_i$ is exactly uniform, including conditionally as in~\eqref{eq:conditional-pa-t}, under its precision-adjusted null. 
\end{proposition}

The proof is in Appendix~\ref{app:exact-inputs}.



\subsection{Conditional Reference Distribution and Pairwise Dependence}
\label{subsec:global-test-validity}

Define the pooled covariance matrix
$\bs{A}^*=\bs A+\bs X\bs X^\top$.
Under the global null, $\bs{A}^* \sim W_d(n+1,\bs\Sigma)$.
The following simple lemma illustrates how conditioning on $\bs{A}^*$ eliminates $\bs\Sigma$ from the conditional distribution of $\bs{X}$ and provides a convenient representation of each pair.

\begin{lemma}[Conditional reference pair]\label{lem:conditional-pair}
Suppose $d\ge2$ and the global null holds.
Then, \begin{equation}\label{eq:reference-pair}
(T_{\mathrm M,i}^2,T_{\mathrm{PA},i}^2)~|~\bs {A}^* \stackrel{d} = \left ( \frac{n\,Z_1^2}{Z_2^2+W+V},
\frac{\nu\,(\sqrt{1- r^2}Z_1+r Z_2)^2}{V} \right ),
\end{equation}
where $r=\widehat{R}_i(\bs A^*)$, and $Z_1, Z_2, V$ and $W$ are all mutually independent, $(Z_1, Z_2) \stackrel{iid} \sim N(0,1)$, $V\sim\chi_\nu^2$, and $W\sim\chi_{d-2}^2$. For $d=2$, $W=0$. In particular, $P_i$ and $Q_i$ are each uniform conditional on $\bs {A}^*$ under the global null.
\end{lemma}

The proof is in Appendix~\ref{app:conditional-representation}. Each pair's conditional distribution depends only on $d$, $n$, and the observable scalar $r$. 
The next result is our key technical contribution, which enables concluding the validity of our two-input Simes combination test through the concavity criterion, which is equivalent to the PDS (Positive Dependence through Stochastic Ordering) condition (\cite{Block1985}) for bivariate copula.

The proof of this lemma requires separate careful arguments for stochastic monotonicity in the two conditioning directions because the statistics have different, dependent random denominators.
When conditioning precision-adjusted evidence on marginal evidence, the resulting angular density is U-shaped for $1\le\nu<2$, so a direct unimodality argument is unavailable; we instead differentiate interval probabilities through their moving endpoints.
The reverse direction requires a different factorization to obtain a symmetric unimodal density, followed by comparisons of intervals that shift away from zero and shrink.

\begin{lemma}[Separate concavity of the pairwise copula]\label{lem:copula-concavity}
For the reference pair~\eqref{eq:reference-pair}, define
$$
P_r^*= \bar {\Psi}_{1,n} \left ( \frac{n\,Z_1^2}{Z_2^2+W+V} \right ),
Q_r^*= \bar{\Psi}_{1, \nu}\left (\frac{\nu\,(\sqrt{1- r^2}Z_1+r Z_2)^2}{V} \right ),
C_r(u,v)=\Pr(P_r^*\le u,Q_r^*\le v).
$$
For $d\ge2$, $\nu\ge1$, and $0\le r\le1$, the function $C_r$ is concave in each argument separately on $[0,1]$.
\end{lemma}

Appendix~\ref{app:dependence} proves this lemma.

\subsection{Conditional Simes Validity and the Global Test}
\label{subsec:wishart-global-test}
The following theorem
providing the validity of our method
in the case of an unknown covariance matrix then follows immediately (see Appendix~\ref{app:global-validity}.
).

\begin{theorem}[Conditional Simes validity with unknown covariance]
\label{thm:global-validity}
In Setting W, with $n\ge d$ and $\bs\mu=\bs0$, the inputs~\eqref{eq:paired-p-values} yield
\begin{equation}\label{eq:pairwise-superuniformity}
\Pr_{\bs0,\bs\Sigma}(S_i\le u\mid\bs {A}^*)
\le u,
\qquad 0\le u\le1,
\qquad i=1,\ldots,d,
\end{equation}
almost surely. Consequently, the global $p$-value~\eqref{eq:ea-global-p-value} satisfies
$\Pr_{\bs0,\bs\Sigma}(P_{\mathrm{EA}}\le\alpha\mid\bs {A}^*)\le\alpha$, $0<\alpha<1$,
for every unknown positive definite $\bs\Sigma$. The same inequality holds unconditionally.
\end{theorem}

\section{Simulation Studies}
\label{sec:simulations}

In this section, we perform simulation studies to compare the performance of our methods with that of several classical methods.

\subsection{Description of Procedures}\label{subsec:simulation-methods}

We compare EA with the procedures below. Let
$\bs X\sim N_d(\bs\mu,\bs\Sigma)$ and
$\bs A\sim W_d(n,\bs\Sigma)$ be independent, with
$n\ge\max(d,5)$, and write
$\bs\Omega=\bs\Sigma^{-1}$ and $\bs S=\bs A/(n+1)$.
Let $\Phi$ and $F_{t_n}$ denote the standard normal and
$t_n$ distribution functions, respectively.
Larger values of each test statistic indicate stronger evidence
against the null. All methods use the Monte Carlo method for
setting critical values described in Section~\ref{sec:mc}.
Feng-COM, QCQ, and ZWS use sample moments in both covariance settings.

\paragraph{Z-Max and Z-TopK \citep{Zhang2017}.}
The Z-Max and Z-TopK methods use the test statistics
$T_{\rm ZM}=w_{[1]}$ and
\[
T_{\rm ZK}
=\max_{1\le k\le K}
\sqrt{\frac{1-k/(n+1)}{2k}}
\left(\sum_{j=1}^k w_{[j]}-k\right),
\]
where $K=\min(40,d,n)$ and
$w_{[1]}\ge\cdots\ge w_{[d]}$ are the ordered squared
precision-adjusted coordinates. Thus, Z-Max uses the largest
squared coordinate, whereas Z-TopK aggregates the largest $k$
squared coordinates and maximizes over $k$.

With known covariance, the adjusted coordinates are
$Z_{{\rm PA},j}=(\bs\Omega\bs X)_j/\sqrt{\Omega_{jj}}$.
With estimated covariance, they are
$Z_{{\rm PA},j}=(\widehat{\bs\Gamma}\bs X)_j/
\sqrt{\widehat\Gamma_{jj}}$,
where $\widehat{\bs\Gamma}$ is obtained by nodewise Lasso.
Specifically, with $\lambda=\sqrt{2\log d/(n+1)}$, define
\[
\begin{aligned}
\widehat{\bs\beta}_j
&=\arg\min_{\bs b}
\left\{
S_{jj}-2\bs S_{j,-j}\bs b
+\bs b^\top\bs S_{-j,-j}\bs b
+2\lambda\|\bs b\|_1
\right\},\\
\widehat\tau_j^2
&=S_{jj}-2\bs S_{j,-j}\widehat{\bs\beta}_j
+\widehat{\bs\beta}_j^\top
 \bs S_{-j,-j}\widehat{\bs\beta}_j
+\lambda\|\widehat{\bs\beta}_j\|_1.
\end{aligned}
\]
Set $\widehat\Gamma_{jj}=1/\widehat\tau_j^2$ and
$\widehat{\bs\Gamma}_{j,-j}
=-\widehat{\bs\beta}_j^\top/\widehat\tau_j^2$.
We use this unsymmetrized estimate and refit it in every replicate.

\paragraph{MinP.}
The MinP method uses the smallest two-sided marginal $p$-value,
with test statistic $T_{\rm MinP}=-\log\min_{1\le j\le d}P_j$.
With known covariance,
$P_j=2\Phi(-|X_j|/\sqrt{\Sigma_{jj}})$;
with estimated covariance,
$P_j=2F_{t_n}(-\sqrt n\,|X_j|/\sqrt{A_{jj}})$.

\paragraph{Quadratic.}
The quadratic method measures the overall departure from zero
using the covariance-adjusted squared norm.
Its test statistic is
$T_{\rm Quad}=\bs X^\top\bs\Omega\bs X$ with known covariance
and $T_{\rm Quad}=n\bs X^\top\bs A^{-1}\bs X$ with estimated
covariance.

\paragraph{Innovated higher criticism (IHC) \citep{HallJin2010}.}
IHC applies higher criticism to transformed coordinates, using
$T_{\rm IHC}
=H(2\Phi(-|\bs V\bs X|);1/d,1/2)/\sqrt{2b-1}$,
where $b= \max(1,\lfloor\log d\rfloor)$ and absolute values
and $\Phi$ act coordinatewise.
To construct $\bs V$, take the Cholesky decomposition
$\bs\Sigma=\bs L\bs L^\top$ and set $\bs U_0=\bs L^{-1}$.
Retain the main diagonal and first $b-1$ subdiagonals of
$\bs U_0$, then normalize each column to Euclidean norm one
to obtain $\bs U$. Set $\bs V=\bs U^\top\bs U_0$.

For ordered inputs $p_{(1)}\le\cdots\le p_{(d)}$, the
higher criticism statistic is
\[
H(\bs p;\ell,u)
=\max_{j:\,\ell\le p_{(j)}\le u}
\frac{\sqrt d\,\{j/d-p_{(j)}\}}
     {\sqrt{p_{(j)}(1-p_{(j)})}},
\]
with value $-\infty$ when the scan is empty.
We evaluate IHC only in the known-covariance setting.

\paragraph{Feng-COM \citep{FengJiangLiLiu2024}.}
Feng-COM combines sum-of-squares and maximum statistics through
$T_{\rm FC}=-\{1-[1-\min(p_S,p_M)]^2\}$.
The component probabilities are $p_S=1-\Phi(T_S)$ and
$p_M=1-G(T_M-2\log d+\log\log d)$, where
$G(x)=\exp\{-e^{-x/2}/\sqrt\pi\}$.
The sum-of-squares statistic is
\[
T_S=
\frac{\bs X^\top\bs D^{-1}\bs X-nd/(n-2)}
     {\sqrt{2\{\operatorname{tr}(\bs R_S^2)-d^2/n\}}},
\]
and the maximum statistic is $T_M=\max_j X_j^2/D_{jj}$,
with $\bs D=\operatorname{diag}(\bs S)$ and
$\bs R_S=\bs D^{-1/2}\bs S\bs D^{-1/2}$.

\paragraph{Observation-level inputs for QCQ and ZWS.}
QCQ and ZWS require individual observations in addition to
$(\bs X,\bs A)$. We construct these using an independent random
orientation. Choose
$\bs H_*\in\mathbb R^{(n+1)\times n}$ satisfying
$\bs H_*^\top\bs H_*=\bs I_n$ and
$\bs H_*^\top\mathbf1=\bs0$.
Independently of $(\bs X,\bs A)$, draw
$\bs U_*\in\mathbb R^{n\times d}$ uniformly from matrices with
orthonormal columns, and set
$\bs C=\bs H_*\bs U_*\bs A^{1/2}$ and
$\bs R_i=\bs C_i+\bs X/\sqrt{n+1}$ for $i=1,\ldots,n+1$.
Here $\bs C_i^\top$ is row $i$ of $\bs C$ and
$\bs A^{1/2}$ is the symmetric positive square root.
Then $\sum_i\bs C_i=\bs0$ and $\bs C^\top\bs C=\bs A$.
Under the Gaussian--Wishart model, the $\bs R_i$ are independent
$N_d(\bs\mu/\sqrt{n+1},\bs\Sigma)$ observations.

\paragraph{QCQ \citep{QiuChenQiu2026}.}
QCQ augments a quadratic statistic with maximum and
higher criticism components. With tuning parameter $M=2$,
its test statistic is
$T_{\rm QCQ}=T_Q+\frac{\sqrt d}{2}(E_\infty+E_H)$,
where $T_Q=\{n\|\bs X\|^2-\operatorname{tr}(\bs A)\}/\sqrt{2J}$.
Writing $t_j=\sqrt n\,X_j/\sqrt{A_{jj}}$ and
$\mathcal H=H(2\Phi(-|\bs t|);\log d/d,d^{-1/4})$,
the enhancement terms are
\[
\begin{aligned}
E_\infty
&=\mathds1\!\left\{
\max_j t_j^2>2\log d+3\log\log d
\right\},\,
E_H
&=\mathds1\!\left\{
\mathcal H>
\sqrt{6\log\log d+\log\log\log d}
\right\}.
\end{aligned}
\]
The variance term uses the reconstructed data
\(
J=\sum_{i\ne j}
\left\{\bs R_i^\top
(\bs R_j-\overline{\bs R}^{(i,j)})\right\}
\left\{\bs R_j^\top
(\bs R_i-\overline{\bs R}^{(i,j)})\right\},
\)
where $\overline{\bs R}^{(i,j)}
=(n-1)^{-1}\sum_{k\ne i,j}\bs R_k$
and the sum defining $J$ is over ordered pairs.
The higher criticism inputs use normal tails in both covariance
settings.

\paragraph{ZWS \citep{ZhangWangShao2025}.}
ZWS combines standardized U-statistics of orders two and six,
using $T_{\rm ZWS}=-\{1-[1-\min(p_2,p_6)]^2\}$.
For $r\in\{2,6\}$, let $p_r=1-\Phi(T_r)$ and
$T_r=(n+1)^{r/2}\mathcal U_r/\sqrt{r!\widehat V_r}$.
The component statistics and variance estimates are
\[
\begin{aligned}
\mathcal U_r
&=\binom{n+1}{r}^{-1}
  \sum_{j=1}^d e_r(R_{1j},\ldots,R_{n+1,j}),\qquad
\widehat V_r
&=\binom{n+1}{r}^{-1}
  \sum_{j,k=1}^d e_r(C_{1j}C_{1k},\ldots,C_{n+1,j}C_{n+1,k}),
\end{aligned}
\]
where $e_r(x_1,\ldots,x_{n+1})
=\sum_{1\le i_1<\cdots<i_r\le n+1}x_{i_1}\cdots x_{i_r}$
is the elementary symmetric polynomial of degree $r$.

\subsection{Monte Carlo Calibration and Uncertainty}\label{sec:mc}

Unless otherwise stated,
we independently create a null bank of size $B = 19,999$ to establish a cutoff for rejecting the null hypothesis for further independent draws. For those independent draws, we generate $R=10,000$ replicates for the Type-1 error and power computations. All methods share the same observations within a replicate, including the same auxiliary orientation.

For the observed score \(T\) and null draws \(T_b^0\), $b=1, \ldots,B$,
calculate $p_B(T)=\frac{1+\sum_{b=1}^B\mathds 1\{T_b^0\ge T\}}{B+1}$, $I=\mathds1\{p_B(T)\le .05\}$.
Consequently, the sample covariance panel compares estimated covariance scores under an oracle setting where the true covariance matrix is used in simulations.

If \(K=\sum_{r=1}^R  I_r\), the rejection probability and its estimated Monte Carlo standard error are
$\widehat\pi=K/R$, $\widehat{\operatorname{MCSE}}(\widehat\pi)=\sqrt{\widehat\pi(1-\widehat\pi)/R}$.
With \(z=\Phi^{-1}(.975)\),
we use the Wilson intervals
$$
 \frac{\widehat\pi+z^2/(2R)\ \pm\ z\sqrt{\widehat\pi(1-\widehat\pi)/R+z^2/(4R^2)}}{1+z^2/R}.
$$

\subsection{Matched Evidence Pairs}\label{subsec:mainsettings}
In our first simulation setting, we consider a pair of ``matched" coordinates, whose effects for the mean and conditional mean have similar magnitudes, but are of the same sign for one coordinate, and of opposite sign for another. Specifically, 
let \(\bs R_b(\rho)=(1-\rho)\bs I_b+\rho\mathbf1_b\mathbf1_b^\top\). Our matched-pair design is
$$
 \bs \Sigma=\operatorname{diag}\{\bs R_2(\rho),\bs I_{d-2}\},\quad
 \bs \mu=a(1,t_*(\rho)+ \eta,0,\ldots,0)^\top,\qquad
 t_*(\rho)=\frac{1-\sqrt{1-\rho^2}}{\rho}.
$$
For \(\eta=0\), the marginal standardized mean vector begins with \((a,at_*)\), and the adjusted standardized mean vector with \((a,-at_*)\). Thus both representations retain the same absolute standardized effects. 
Further, 
the correlation between marginal and adjusted statistics at the same coordinate is controlled by $\rho$ to be \(\sqrt{1-\rho^2}\). 
This gives EA an opportunity to combine supporting evidence locally. 

Discordant behavior of strongly associated signals when adjusted for precision can be encountered in empirical settings.
For instance, 
when 
jointly measuring a protein and its phosphorylated form \citep{Wu2011phosphorylation}, shared variation in protein abundance can induce strong positive dependence, while a biological perturbation changes protein abundance and phosphorylation unequally. 
The phosphorylated form may increase less than expected given the accompanying increase in total protein. 
The matched-pair construction provides an idealized representation of this setting.

Table~\ref{tab:designs} gives nine matched settings (M1--M9) that we consider. 
To show evidence occurring across different groups, B1--B4 instead use
$$
 \bs \Sigma=\operatorname{diag}\{\bs R_2(.95),\bs R_2(.9),\bs I_{d-4}\},\qquad
 \bs \mu=(3.5,3.5,c,-c,0,\ldots,0)^\top.
$$
B1/B2 use \((d,n)=(50,249)\), B3/B4 use \((100,499)\); within each pair \(c \in \{.6,.8\} \). Three controls are retained: C1 has the M5 covariance with \(\bs \mu=(3.5,3.5,0,\ldots,0)^\top\); C2 has \(d=50,n=249,\bs \Sigma = \bs I_d,\bs \mu=3.5 \bs e_1\); C3 has the M5 covariance with \(\bs \mu=1.25 \bs e_1\). 
These controls place the results in context by showing how EA’s relative performance varies.

We consider a broader set of mean and covariance settings in \ref{sec:app-broad-study}. In general, our methods are competitive with the baselines across settings \footnote{ The reproducible codes for the simulations and data analysis are available at \url{https://github.com/deepraghosh/evidence-adaptive-test}}.

\begin{table}\centering
\caption{Fixed matched-pair settings, all using the covariance and mean from Section~\ref{subsec:mainsettings}.}\label{tab:designs}
{\small\setlength{\tabcolsep}{9pt}\renewcommand{\arraystretch}{1.12}
\begin{tabular}{lrrrrr}
\toprule
Setting & $d$ & $n$ & $\rho$ & $a$ & $\eta$ \\
\midrule
M1 & 20 & 99 & 0.9 & 3.5 & 0 \\
M2 & 20 & 99 & 0.95 & 3.5 & 0 \\
M3 & 50 & 249 & 0.9 & 3.5 & 0 \\
M4 & 50 & 99 & 0.95 & 3.5 & 0 \\
M5 & 50 & 249 & 0.95 & 3.5 & 0 \\
M6 & 50 & 249 & 0.95 & 3.5 & -0.05 \\
M7 & 50 & 249 & 0.95 & 3.5 & +0.05 \\
M8 & 50 & 499 & 0.95 & 3.5 & 0 \\
M9 & 100 & 499 & 0.95 & 3.75 & 0 \\
\bottomrule
\end{tabular}}
\end{table}

Tables~\ref{tab:main-known} and~\ref{tab:main-estimated} split the comparisons by covariance information. Abbreviations are ZM: Z-Max; ZK: Z-TopK; FC: Feng-COM; MP: MinP; Q: Quadratic. The ZWS column is the adaptive combination of orders two and six. IHC is present only with known covariance.

\begin{table}\centering
\caption{Simulated power in fixed designs under known-covariance settings.}\label{tab:main-known}
{\small\setlength{\tabcolsep}{3pt}\renewcommand{\arraystretch}{1.12}
\begin{tabular}{lrrrrrrrrr}
\toprule
Setting & EA & ZM & ZK & IHC & FC & QCQ & ZWS & MP & Q \\
\midrule
M1 & \textbf{0.795} & 0.698 & 0.687 & 0.192 & 0.641 & 0.646 & 0.626 & 0.703 & 0.689 \\
M2 & \textbf{0.815} & 0.706 & 0.710 & 0.196 & 0.653 & 0.665 & 0.654 & 0.699 & 0.742 \\
M3 & \textbf{0.710} & 0.611 & 0.596 & 0.178 & 0.555 & 0.539 & 0.527 & 0.603 & 0.467 \\
M4 & \textbf{0.727} & 0.601 & 0.611 & 0.184 & 0.559 & 0.531 & 0.541 & 0.607 & 0.509 \\
M5 & \textbf{0.722} & 0.600 & 0.605 & 0.188 & 0.564 & 0.551 & 0.538 & 0.603 & 0.507 \\
M6 & \textbf{0.825} & 0.784 & 0.787 & 0.208 & 0.557 & 0.545 & 0.529 & 0.603 & 0.599 \\
M7 & \textbf{0.635} & 0.405 & 0.415 & 0.156 & 0.574 & 0.556 & 0.553 & 0.603 & 0.431 \\
M8 & \textbf{0.728} & 0.602 & 0.605 & 0.183 & 0.567 & 0.546 & 0.534 & 0.605 & 0.512 \\
M9 & \textbf{0.754} & 0.631 & 0.610 & 0.180 & 0.571 & 0.563 & 0.549 & 0.625 & 0.404 \\
B1 & \textbf{0.689} & 0.346 & 0.438 & 0.179 & 0.656 & 0.617 & 0.656 & 0.661 & 0.535 \\
B2 & \textbf{0.823} & 0.667 & 0.748 & 0.214 & 0.662 & 0.623 & 0.664 & 0.663 & 0.685 \\
B3 & \textbf{0.597} & 0.280 & 0.371 & 0.174 & 0.569 & 0.554 & 0.577 & 0.580 & 0.383 \\
B4 & \textbf{0.748} & 0.590 & 0.677 & 0.220 & 0.575 & 0.558 & 0.582 & 0.581 & 0.510 \\
C1 & 0.584 & 0.0574 & 0.0597 & 0.0518 & 0.652 & 0.612 & 0.650 & \textbf{0.657} & 0.326 \\
C2 & \textbf{0.617} & \textbf{0.617} & 0.584 & 0.112 & 0.549 & 0.515 & 0.517 & \textbf{0.617} & 0.327 \\
C3 & 0.735 & 0.793 & \textbf{0.859} & 0.226 & 0.0758 & 0.0742 & 0.0728 & 0.0697 & 0.427 \\
\bottomrule
\end{tabular}}
\end{table}

\begin{table}\centering
\caption{Simulated power in fixed designs under sample covariance settings.}\label{tab:main-estimated}
{\small\setlength{\tabcolsep}{3pt}\renewcommand{\arraystretch}{1.12}
\begin{tabular}{lrrrrrrrr}
\toprule
Setting & EA & ZM & ZK & FC & QCQ & ZWS & MP & Q \\
\midrule
M1 & \textbf{0.733} & 0.604 & 0.593 & 0.641 & 0.646 & 0.626 & 0.683 & 0.591 \\
M2 & \textbf{0.750} & 0.482 & 0.486 & 0.653 & 0.665 & 0.654 & 0.670 & 0.625 \\
M3 & \textbf{0.658} & 0.492 & 0.467 & 0.555 & 0.539 & 0.527 & 0.588 & 0.388 \\
M4 & 0.552 & 0.344 & 0.341 & 0.559 & 0.531 & 0.541 & \textbf{0.570} & 0.299 \\
M5 & \textbf{0.672} & 0.355 & 0.349 & 0.564 & 0.551 & 0.538 & 0.587 & 0.427 \\
M6 & \textbf{0.751} & 0.488 & 0.464 & 0.557 & 0.545 & 0.529 & 0.587 & 0.508 \\
M7 & \textbf{0.605} & 0.249 & 0.248 & 0.574 & 0.556 & 0.553 & 0.587 & 0.364 \\
M8 & \textbf{0.709} & 0.400 & 0.374 & 0.567 & 0.546 & 0.534 & 0.599 & 0.471 \\
M9 & \textbf{0.694} & 0.398 & 0.331 & 0.571 & 0.563 & 0.549 & 0.613 & 0.342 \\
B1 & \textbf{0.656} & 0.106 & 0.173 & 0.656 & 0.617 & 0.656 & 0.644 & 0.459 \\
B2 & \textbf{0.759} & 0.239 & 0.361 & 0.662 & 0.623 & 0.664 & 0.646 & 0.600 \\
B3 & \textbf{0.578} & 0.0975 & 0.166 & 0.569 & 0.554 & 0.577 & 0.573 & 0.314 \\
B4 & \textbf{0.686} & 0.236 & 0.346 & 0.575 & 0.558 & 0.582 & 0.573 & 0.426 \\
C1 & 0.585 & 0.0708 & 0.0890 & \textbf{0.652} & 0.612 & 0.650 & 0.645 & 0.270 \\
C2 & 0.574 & \textbf{0.601} & 0.572 & 0.549 & 0.515 & 0.517 & 0.600 & 0.283 \\
C3 & \textbf{0.598} & 0.170 & 0.227 & 0.0758 & 0.0742 & 0.0728 & 0.0711 & 0.350 \\
\bottomrule
\end{tabular}}
\end{table}

In the designs reported in Tables~\ref{tab:main-known} and~\ref{tab:main-estimated}, EA with known covariance has more power than the baselines in 13 of the 16 settings, including all nine matched pairs. In the sample covariance panel, results are similar. EA improves on both Z procedures in 15 of the 16 sample covariance alternatives. These designs do not isolate out a single attributable signal, which is where Z-Max excels in detectability. 

Some other results where competing methods perform well are as follows. In M4, with only \(n=99\) Wishart degrees at \(d=50\), sample covariance EA has power $.552$ and performs poorly compared to MinP ($.570$), although it improves on both Z methods. In C1, coherent signal favors marginal and moment methods: known-covariance MinP has power $.657$ compared with EA $.584$. The plausible reason can be understood by realizing that the precision-adjusted effect for the $(3.5,3.5)$ duo is $(.56, .56)$ which means the marginal evidence dominates mostly incurring a calibration cost that adds no extra evidence, whereas the MinP solely focuses on the effect standing out. In C2, known-covariance EA, Z-Max and MinP coincide at $.617$, while estimated-covariance EA ($.574$) is inferior to Z-Max ($.601$). In C3, the known-covariance isolated signal favors Z-TopK ($.859$) and Z-Max ($.793$) over EA ($.735$). This control design, in the known covariance case, translates to a situation where the marginal evidence of $(1.25,0)$ lacks in strength to the adjusted evidence of $(4.003, -3.803)$. Z-Max acts more sensitive to strong precision-adjusted evidence and Z-TopK exploits both large squared adjusted effects. EA drops out in power since the marginals do not retain considerable information. A broader study is presented in \hyperlink{sec:supp}{S2}.

\subsection{Unequal Responses in a Correlated Module}
\label{subsec:correlated-module}

In our second setting, we consider a small group of positively correlated measurements whose means change in the same direction, but by different amounts. Such configurations are relevant to gene-set analyses, where measurements in a biological pathway can share variation while responding unequally to a perturbation. \citet{ShiEtAl2011} considered Gaussian signal genes with heterogeneous mean effects in a compound-symmetric block, together with independent null genes. 
Similar models also appear in the simulations of \citet[Section~4.1]{HuTongGenton2024}. 
Motivated by these models, we consider a setting in which one coordinate responds more strongly than the other coordinates in its group.

Specifically, using $\bs R_b(\rho)$ from Section~\ref{subsec:mainsettings}, let
$$
\bs\Sigma=\operatorname{diag}\{\bs R_5(.75),\bs I_{d-5}\},
\qquad
\bs\mu=a(1,1/2,1/2,1/2,1/2,0,\ldots,0)^\top.
$$
We take $(d,n)=(50,499)$ and $a\in\{3,3.5,4\}$. The covariance has the representation $X_j=\mu_j+\sqrt{.75}F+.5\varepsilon_j$ for $j\le5$, where $F,\varepsilon_1,\ldots,\varepsilon_5$ are independent standard normal variables. Thus the active measurements have a shared source of variation and different mean responses. 
In the Gaussian sample mean realization with $500$ observations, $a=3.5$ corresponds to individual-level standardized mean shifts of approximately $.157$ for the leading coordinate and $.0783$ for the other four. 

The first five standardized marginal effects are $(a,a/2,a/2,a/2,a/2)$. Precision adjustment gives $a(7,-1,-1,-1,-1)/(2\sqrt{13})$, since the inverse block is $\bs R_5(.75)^{-1}=4\bs I_5-(3/4)\mathbf1_5\mathbf1_5^\top$. The leading coordinate has standardized effects $a$ and approximately $.971a$ in the two representations, 
only mildly correlated with correlation coefficient $2/\sqrt{13}\approx.555$. 
The four smaller marginal effects become weak adjusted effects of approximately $-.139a$. Hence the two representations supply complementary evidence without the coordinatewise matching of absolute effects in Section~\ref{subsec:mainsettings}. The comparisons concern the critical values set via Monte Carlo; in Setting W this retains the oracle null critical values 
using the true covariance, as in Section~\ref{sec:mc}.

\begin{table}\centering
\caption{Simulated power for unequal responses in a correlated module. Bold denotes the largest estimated power in each row. IHC is evaluated only with known covariance.}\label{tab:new_module-power}
{
\begin{tabular}{lrrrrrrrrr}
\toprule
$a$ & EA & ZM & ZK & IHC & FC & QCQ & ZWS & MP & Q \\
\midrule
\multicolumn{10}{l}{\emph{Known covariance}}\\
3 & \textbf{0.473} & 0.386 & 0.366 & 0.113 & 0.416 & 0.391 & 0.411 & 0.421 & 0.280 \\
3.5 & \textbf{0.671} & 0.565 & 0.530 & 0.124 & 0.576 & 0.546 & 0.569 & 0.603 & 0.400 \\
4 & \textbf{0.834} & 0.736 & 0.695 & 0.132 & 0.727 & 0.694 & 0.714 & 0.771 & 0.544 \\
\midrule
\multicolumn{10}{l}{\emph{Estimated covariance}}\\
3 & \textbf{0.464} & 0.351 & 0.327 & --- & 0.416 & 0.391 & 0.411 & 0.420 & 0.260 \\
3.5 & \textbf{0.658} & 0.546 & 0.503 & --- & 0.576 & 0.546 & 0.569 & 0.601 & 0.369 \\
4 & \textbf{0.818} & 0.733 & 0.687 & --- & 0.727 & 0.694 & 0.714 & 0.767 & 0.510 \\
\bottomrule
\end{tabular}}
\end{table}

Table~\ref{tab:new_module-power} shows that EA has higher estimated power than compared methods
at all three amplitudes, under both known and estimated covariance. At $a=3.5$, its powers are $0.677$ and $0.666$, compared with $0.609$ for MinP and $0.603$ for MinP, the strongest respective comparators. Across the six comparisons with the strongest baseline, the improvement ranges from $4.93$ to $6.71$ percentage points.

\subsection{Localized Peaks with Moving Average Noise}
\label{subsec:moving average}

Our third setting concerns a localized mean change in measurements with short-range dependence. Local peaks and moving average preprocessing arise in mass spectrometry data analysis \citep{YangHeYu2009}. Overlapping averages share noise contributions even when the underlying errors are independent.
Similar models are also used in high-dimensional testing simulations by \citet[Section~5.1]{ChenQin2010}. 

We consider such a model class with a three-coordinate peak and first-order moving average noise.
Let $\varepsilon_0,\ldots,\varepsilon_d$ be independent standard normal variables and set
$$
X_j=\mu_j+\frac{5\varepsilon_j+4\varepsilon_{j-1}}{\sqrt{41}},
\qquad j=1,\ldots,d.
$$
The covariance matrix has unit diagonal, adjacent entries $20/41$, and all other entries zero. Thus adjacent measurements have correlation approximately $.488$, while nonadjacent measurements are independent. We take $(d,n)=(50,499)$ and $\bs\mu=a\{\bs e_{26}+(\bs e_{25}+\bs e_{27})/3\}$, with $a\in\{3,3.5,4\}$. The central response is three times the response at each of its two neighbors; all other means vanish.

At the central coordinate, the standardized marginal effect is $a$. Direct inversion of the stated finite covariance matrix gives an adjusted standardized effect of approximately $.996a$ and correlation approximately $.469$ between its marginal and adjusted statistics. The central effect is therefore retained in both representations.
Precision adjustment also produces alternating signed effects beyond the original three-coordinate support. 
EA can combine complementary evidence at the central peak without using its location or shape as input to the test.

\begin{table}\centering
\caption{Simulated power for localized peaks with moving average noise. Bold denotes the largest estimated power in each row. IHC is evaluated only with known covariance.}\label{tab:new_ma-power}
{
\begin{tabular}{lrrrrrrrrr}
\toprule
$a$ & EA & ZM & ZK & IHC & FC & QCQ & ZWS & MP & Q \\
\midrule
\multicolumn{10}{l}{\emph{Known covariance}}\\
3 & \textbf{0.514} & 0.456 & 0.336 & 0.171 & 0.371 & 0.255 & 0.340 & 0.423 & 0.312 \\
3.5 & \textbf{0.717} & 0.637 & 0.480 & 0.214 & 0.538 & 0.368 & 0.501 & 0.606 & 0.448 \\
4 & \textbf{0.871} & 0.799 & 0.633 & 0.264 & 0.715 & 0.504 & 0.673 & 0.781 & 0.601 \\
\midrule
\multicolumn{10}{l}{\emph{Estimated covariance}}\\
3 & \textbf{0.491} & 0.446 & 0.402 & --- & 0.371 & 0.255 & 0.340 & 0.420 & 0.289 \\
3.5 & \textbf{0.686} & 0.646 & 0.585 & --- & 0.538 & 0.368 & 0.501 & 0.601 & 0.418 \\
4 & \textbf{0.848} & 0.810 & 0.764 & --- & 0.715 & 0.504 & 0.673 & 0.778 & 0.558 \\
\bottomrule
\end{tabular}}
\end{table}

Table~\ref{tab:new_ma-power} shows higher estimated power for EA than for
the compared methods in both covariance settings. At $a=3.5$, 
the powers are $0.722$ and $0.689$, compared with $0.638$ for Z-Max and $0.646$ for Z-Max, the strongest respective comparators. 
Across the six comparisons, the improvement with respect to the best baselines ranges from $3.47$ to $8.37$ percentage points. Additional  results appear in \ref{sec:app-new-settings-check}.

\subsection{Matched Pair Robustness} \label{subsec:curves}
\begin{figure}
\centering
\begin{minipage}{0.32\textwidth}
\centering

\includegraphics[width=\linewidth]
{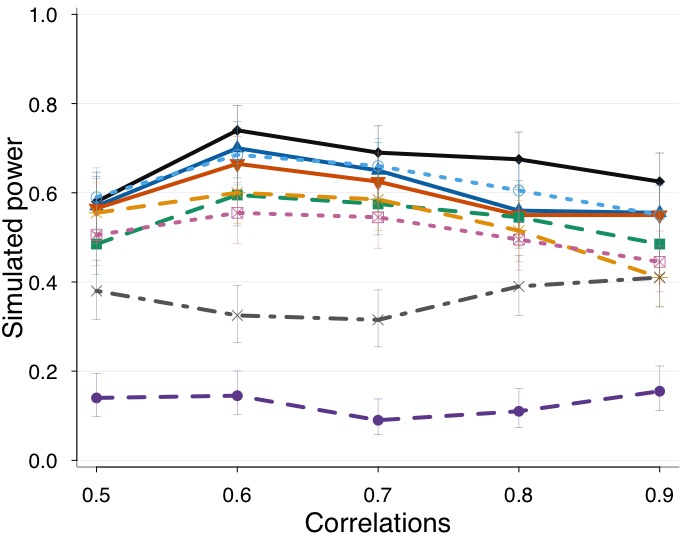}
\end{minipage}
\hfill
\begin{minipage}{0.32\textwidth}
\centering

\includegraphics[width=\linewidth]
{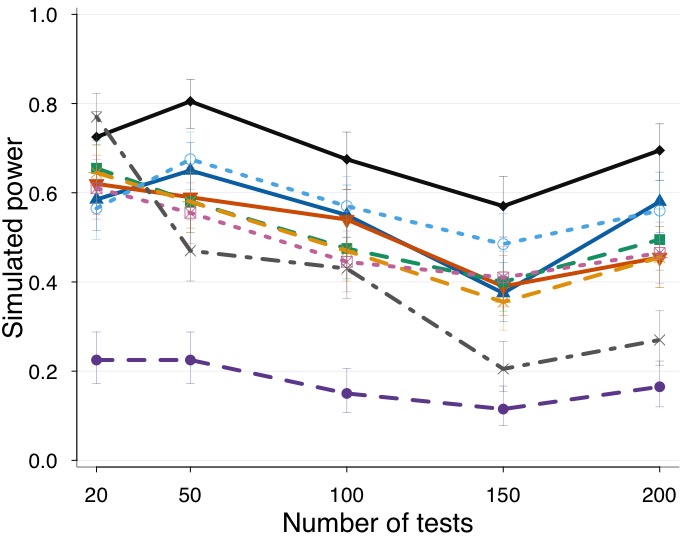}
\end{minipage}
\hfill
\begin{minipage}{0.32\textwidth}
\centering

\includegraphics[width=\linewidth]
{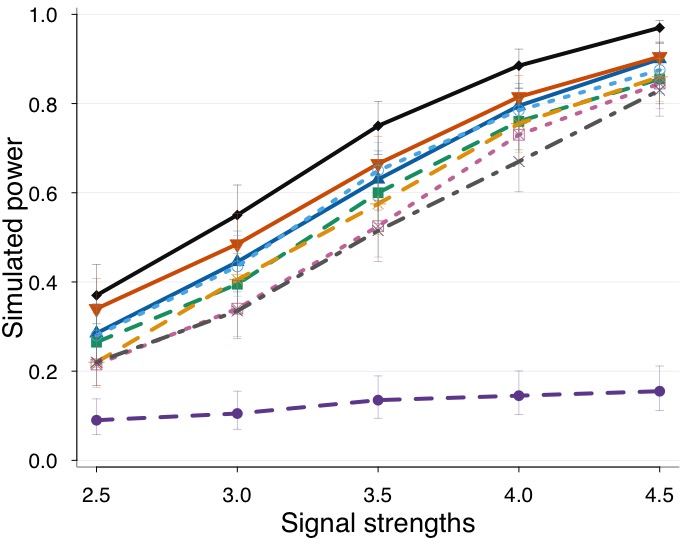}
\end{minipage}

\vspace{1em}

\begin{minipage}{0.32\textwidth}
\centering

\includegraphics[width=\linewidth]
{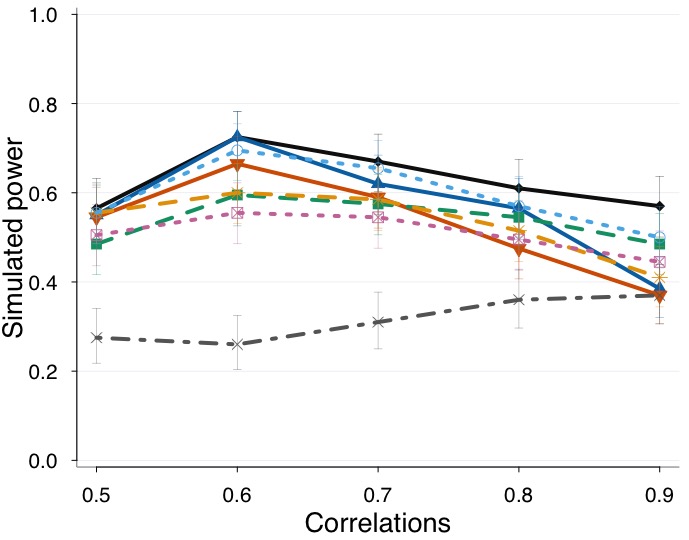}
\end{minipage}
\hfill
\begin{minipage}{0.32\textwidth}
\centering
\includegraphics[width=\linewidth]
{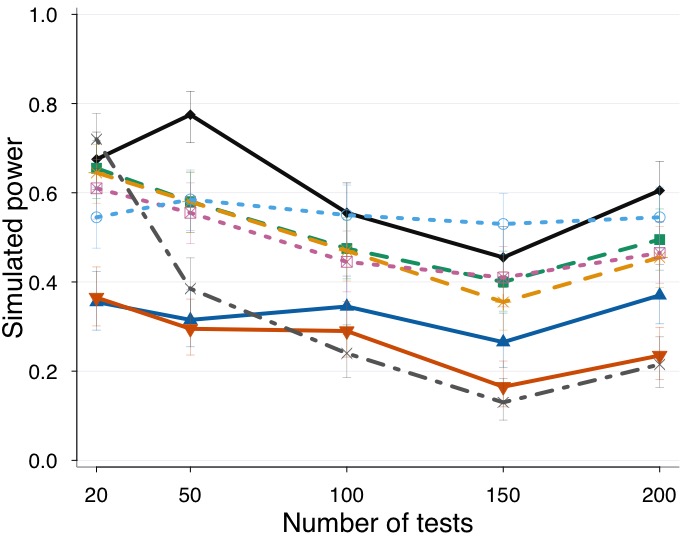}
\end{minipage}
\hfill
\begin{minipage}{0.32\textwidth}
\centering
\includegraphics[width=\linewidth]
{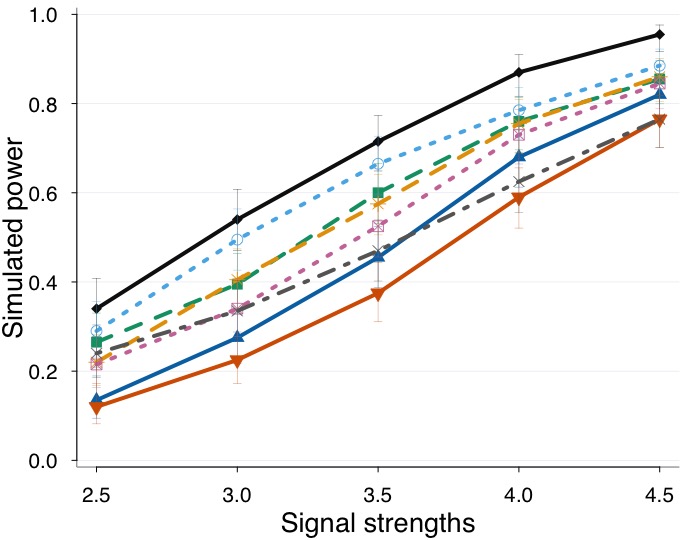}
\end{minipage}

\caption{Simulated power in known-covariance (top row) and sample covariance (bottom row) settings in the matched-pair designs. From left to right, the horizontal axes vary in correlation, dimension and amplitude. Whiskers are pointwise 95\% Wilson intervals; $B=1999$, $R=2000$, $\alpha=.05$. Methods compared are EA (filled diamond, black); Z-Max (filled upward triangle, blue); Z-TopK (filled downward triangle, vermilion); IHC (filled circle, purple); Feng-COM (filled square, green); QCQ (asterisk, orange); ZWS (crossed square, reddish purple); MinP (open circle, sky blue); Quadratic (cross, gray). }
\label{fig:main-text}
\end{figure}

Figure~\ref{fig:main-text} shows an additional power study, under three types of dynamic designs within the matched pair framework. The correlation study fixes \(d=50, n=249, a=3.5, \eta=0\) and varies \(\rho \in \{.5, .6, .7, .8, .9\}\). The dimension study fixes \(\rho=.95,a=3.5,\eta=0\), varies \(d \in \{20, 50, 100, 150, 200\}\), and sets \(n = 5d-1\). The amplitude design fixes \(d=50,n=249,\rho=.95,\eta=0\) and varies \(a \in \{2.5,3,3.5,4,4.5\}\). These extend the scope of the EA test from the perspective of scalability and dependency. The EA test retains improved power. 
The null rejection intervals mostly contain $\alpha$, see \ref{sec:app-null-rejection-intervals}.

\section{Data Analysis Illustration}
\label{sec:real data analysis}


The Cancer Genome Atlas (TCGA) was established to characterize the molecular architecture of human cancers through coordinated measurements of genomic, transcriptomic, epigenetic, and proteomic alterations. TCGA produced data suitable both for cancer-specific investigations and for comparative analyses across disease sites \citep{TCGA2013PanCancer}.

In this paper, as an illustation, we  analyze data from  protein expression measurements obtained by reverse-phase protein array (RPPA). RPPA assays quantify the relative abundance or activation state of selected proteins and phosphoproteins, including components of signaling, proliferation, apoptosis, and DNA-repair pathways. The public PanCancer Atlas versions of the RPPA and clinical data were obtained from the cBioPortal DataHub \citep{Gao2013cBioPortal}. The RPPA profiles contain processed log2-scale measurements \footnote{ The data are publicly available on the cBioPortal DataHub TCGA COAD/READ PanCancer Atlas study, at \url{https://github.com/cBioPortal/datahub/tree/master/public/coadread_tcga_pan_can_atlas_2018}}.

RPPA panels provide a natural setting for this comparison because proteins and phosphoproteins belonging to shared signaling and regulatory pathways can exhibit substantial dependence.
In contrast, 
 biological differences between tumor groups may be concentrated in a relatively small subset of the measured features.
 
 We looked at two such settings:
 a colorectal cancer study representing a high-dimensional
profile dominated by strong marginal signals in relatively few coordinates, 
and a  stomach adenocarcinoma (STAD) analysis that
examines whether conditioning  on correlated proteins can strengthen effects. 
In each illustration, the multivariate response is the vector of RPPA measurements from a single primary tumor, while the clinical or molecular classification defines the two populations being compared. The proposed procedure was compared 
with baselines introduced in the simulation sections by applying to the standardized two-sample mean contrast and pooled within-group covariance matrix. 

The colorectal illustration is based on the TCGA colon and rectal adenocarcinoma PanCancer Atlas study.  TCGA provided  molecular characterization of colon and rectal carcinomas and documented their principal genomic alterations \citep{TCGA2012COADREAD}. A later comparative analysis of gastrointestinal adenocarcinomas defined molecular classes that included chromosomal-instability and genome-stable colorectal tumors \citep{Liu2018GIAC}. 

The present analysis focuses on colon adenocarcinomas classified as chromosomal instability (CIN) or genomically stable (GS). The CIN tumors have substantial changes in the amount of DNA carried on chromosomes; for instance, extra or missing copies of chromosomes.
GS tumors have relatively few of those large-scale chromosome changes. 
The colorectal RPPA matrix contained 464 tumor samples and 198 protein or phosphoprotein features. After matching the molecular measurements to the clinical annotations, the analysis included 175 CIN tumors and 45 GS tumors. The final multivariate sample contained 189 RPPA features. 




The question of interest is whether CIN and GS colon tumors have the same mean protein expression profile.
If $\bs{\mu}_{\mathrm{CIN}}$ and $\bs{\mu}_{\mathrm{GS}}$ denote the corresponding population mean vectors, the global null hypothesis is $
H_0:\bs{\mu}_{\mathrm{CIN}}=\bs{\mu}_{\mathrm{GS}}$.

The second illustration uses the TCGA stomach adenocarcinoma (STAD) PanCancer Atlas RPPA data \citep{TCGA2014STAD} to compare protein expression profiles by pathological lymph node status. 
Tumors classified as N0 constitute the node-negative group (no indication of regional lymph-node metastasis), while N1, N2, N3, N3A, and N3B tumors constitute the node-positive group (various stages of metastasis). 
Tumors with unknown or uninterpretable nodal status are excluded.
The analysis includes 108 node-negative and 230 node-positive tumors measured on 193 RPPA features.
This analysis asks whether the mean multivariate RPPA profile is the same in node-negative and node-positive gastric adenocarcinomas. Writing $\bs{\mu}_{N0}$ and $\bs{\mu}_{N+}$ for the corresponding mean vectors, the null hypothesis is $H_0:\bs{\mu}_{N0}=\bs{\mu}_{N+}$.




\begin{table}
\centering
\caption{TCGA illustration: $p$-values of the various methods.}
\label{tab:data-pvalues}
\small
\setlength{\tabcolsep}{3pt}
\begin{tabular}{@{}lccccccc@{}}
\toprule
Dataset & EA & Z-Max & Feng-COM & QCQ$^\dagger$
& ZWS-$\{2,6\}$ & MinP & Hotelling \\
\midrule
Colorectal
& $2.31\times10^{-9}$
& $2.33\times10^{-10}$
& $2.62\times10^{-10}$
& $1.17\times10^{-77}$
& $4.33\times10^{-5597}$
& $1.15\times10^{-9}$
& 0.0961 \\
\addlinespace
STAD
& $2.64\times10^{-4}$
& 0.0442
& 0.284
& 0.410
& 0.311
& 0.170
& 0.0129 \\
\bottomrule
\end{tabular}
\end{table}

We present the results from both the studies in Table \ref{tab:data-pvalues}.
Here, Z-Max uses pooled coordinate standardization and an asymptotic Gumbel null distribution. 
Feng-COM is calculated directly from the group-mean contrast and pooled within-group covariance matrix, whereas QCQ and ZWS use the original
observations in their two groups. 
Feng-COM combines a normally calibrated sum component and a maximum component with Gumbell reference null. QCQ and the order-specific ZWS statistics use asymptotic standard normal null approximations; the ZWS global $p$-value combines the order-two and order-six $p$-values. 

In the colorectal data example,
our method detects a global difference alongside the marginal combination procedures when Hotelling’s test does not reject.
In the STAD analysis, it continues to detect a difference when methods based on asymptotic distributional properties do not.
The Hotelling and Z-Max tests also reject the null. 

\section{Concluding Remarks}\label{sec:discussion}
This paper introduces a covariance-informed approach to global Gaussian mean testing that combines marginal and conditional evidence through coordinatewise Simes recalibration and Bonferroni correction. Its key methodological innovation lies in exploiting complementary signal representations to improve power while retaining simple multiplicity adjustment. A principal theoretical contribution is finite-sample Type-I error control under both known and independently estimated covariance, without restrictions on correlation signs. Particularly novel is the estimated-covariance analysis, which establishes the validity of combining studentized statistics with different degrees of freedom through a covariance-free conditional representation and copula concavity.

Extensive simulations show that the proposed EA test is generally competitive with existing global tests and can achieve substantial power gains under structured alternatives. Notable improvements arise when two coordinates exhibit comparable marginal and conditional effects, with concordant signs at one coordinate and discordant signs at the other. The three control designs, when compared with matched pair designs, highlight EA can benefit from two informative, imperfectly redundant sources of evidence, unlike strong marginal effects, strong adjusted effects or identical uncorrelated effects.The correlated module including more than a matched pair of effects show that EA can be useful when a leading response sits within a correlated group whose members have weaker responses. 
The moving-average design extends the study to a local signal peak embedded in a connected sequence of dependent measurements, where dependence continues throughout the coordinate sequence, and precision adjustment produces alternating effects extending beyond the marginal signal support. An application to TCGA carcinoma data illustrates the method's practical potential where simulation-conforming design assessment is beyond the scope.

Beyond global testing, the framework provides componentwise inference for $H_{i,0}:\mu_i=\delta_i=0$, where $\bs\delta=\bs\Sigma^{-1}\bs\mu$, with strong familywise error rate control under known covariance and weak control under estimated covariance. It also suggests extensions to FDR-controlled discovery through the PTDN framework \citep{GhoshSarkar2025} and shifted-BH methodology \citep{SarkarZhang2025,GhoshSarkar2026}. Extending these guarantees to estimated covariance and regularized precision estimation remains an important direction for future research. More broadly, this work demonstrates how complementary covariance-informed evidence can enhance testing power without sacrificing the simplicity of rigorous error control.

{\bf AI Usage.}
The authors used OpenAI GPT-6.0 Sol for tasks such as exploring proof strategies, improving the presentation of certain mathematical arguments, and cleaning up the simulation code. 
The research ideas, methodology, and principal theoretical contributions were developed by the authors, who reviewed and verified all AI-assisted material and take full responsibility for the final manuscript.

\section*{Acknowledgements}
The research of SKS is supported by NSF Grant-DMS 2210687.

\appendix
\section{Proofs for the Independent-Wishart Construction}\label{app:proofs}

\subsection{Validity of the Two Inputs}
\label{app:exact-inputs}

\begin{proof}[Proof of Proposition~\ref{prop:exact-inputs}]
The marginal statement follows from $A_{ii}/\Sigma_{ii}\sim\chi_n^2$ and independence of $\bs X$ and $\bs A$. For the second statement, we first note that
\begin{equation}\label{eq:wishart-regression-law}
\widehat{\bs\beta}_i\mid\bs A_{-i,-i}
\sim N_{d-1}\!\left(\bs\beta_i,
 v_i\bs A_{-i,-i}^{-1}\right),
\qquad D_i/v_i\sim\chi_\nu^2,
\end{equation}
with $D_i$ independent of $(\widehat{\bs\beta}_i,\bs A_{-i,-i})$; see, for instance, \cite{Anderson2003}.
Combining \eqref{eq:wishart-regression-law} with~\eqref{eq:precision-residual} yields
$$
X_i-\widehat{\bs\beta}_i^\top\bs X_{-i}
\mid(\bs X_{-i},\bs A_{-i,-i})
\sim N\!\left(v_i\delta_i,\,v_i(1+h_i)\right),
$$
independently of $D_i$. Setting $\delta_i=0$ gives~\eqref{eq:conditional-pa-t}.
\end{proof}

\subsection{Conditional Distribution and Reference-Pair Representation}
\label{app:conditional-representation}

\begin{proof}[Proof of Lemma~\ref{lem:conditional-pair}]
Under the global null the joint density of $(\bs X,\bs A)$ is proportional to
$$|\bs A|^{(n-d-1)/2}|\bs\Sigma|^{-(n+1)/2}
\textrm{exp}\left[-\frac{1}{2}\textrm{tr}\{\bs\Sigma^{-1}
(\bs A+\bs X\bs X^\top)\}\right].
$$
Using the change of variables, $\bs{A} \rightarrow \bs {A}^*=\bs A+\bs X\bs X^\top$, and that $|\bs {A}^* -\bs X \bs X^\top|
=|\bs {A}^* |(1-\bs X^\top{\bs {A}^*}^{-1}\bs X)$, where $\bs X^\top {\bs {A}^*}^{-1}\bs X<1$, since $\bs{A}$ is positive definite, we have the following conditional density of $\bs{X}$ given $\bs{A}^*$:
$$
f(\bs x\mid\bs {A}^*)
\propto(1-\bs x^\top{\bs {A}^*}^{-1}\bs x)^{\nu/2-1}
\mathds{1}\{\bs x^\top{\bs{A}^*}^{-1}\bs x<1\}.
$$
Thus, using the symmetric positive definite square root of $\bs{A}^*$, we have
$$
\bs U ={\bs{A}^*}^{-1/2}\bs X
\ \overset{d}{=}\
\frac{\bs Z}{\sqrt{\|\bs Z\|^2+V}}
\quad\text{conditionally on }\bs {A}^*,
$$
where $\bs Z\sim N_d(\bs 0,\bs I)$ and
$V \sim\chi_\nu^2$ independently.

Now, defining the unit vectors
$
\bs{a}_i=\frac{{\bs{A}^*}^{1/2}\bs {e}_i}{\sqrt{{A^*}_{ii}}}, \quad \bs{b}_i=\frac{{\bs{A}^*}^{-1/2}\bs {e}_i} {\sqrt{({\bs {A}^*}^{-1})_{ii}}},
$
we have from $A_{ii} = A^*_{ii}- X_i^2$ that
$$ T_{\mathrm M,i}^2 = \frac{n\,(\bs {a}_i^\top\bs U)^2}{1-(\bs a_i^\top\bs U)^2} \overset{d}{=} \frac{n\,(\bs {a}_i^\top\bs Z)^2}{\|\bs{Z}\|^2+ V - (\bs{a}_i^\top\bs Z)^2}. $$ 
Also, by simple algebra,
$$
\begin{aligned}
\frac{1}{\nu} T_{\mathrm{PA},i}^2 &= \frac{\bs{X}^\top \bs{A}^{-1}\bs{X} - \bs{X}_{-i}^\top \bs{A}_{-i,-i}^{-1}\bs{X}_{-i}}{1+\bs{X}_{-i}^\top \bs{A}_{-i,-i}^{-1}\bs{X}_{-i}} =  \frac{1+\bs{X}^\top \bs{A}^{-1}\bs{X}}{1+\bs{X}_{-i}^\top \bs{A}_{-i,-i}^{-1}\bs{X}_{-i}} - 1  \\ &= \frac{|\bs{A}+\bs{X}\bs{X}^\top|}{|\bs{A}_{-i,-i}+\bs{X}_{-i}\bs{X}_{-i}^\top|} \div
\frac{|\bs{A}|}{|\bs{A}_{-i,-i}|} - 1 = \frac{(\bs{A}^{-1})_{ii}}{((\bs{A}+\bs{X}\bs{X}^\top)^{-1})_{ii}} - 1.
\end{aligned}
$$
Hence
$$
\begin{aligned}
\frac{1}{\nu}T_{\textrm{PA},i}^2 &= \frac{\bs{e}_i^\top\left(\bs{A}^* -
\bs{X}\bs{X}^\top \right )^{-1}\bs{e}_i}{\left({\bs{A}^*}^{-1} \right)_{ii}} - 1 =  \frac{\bs{e}_i^\top{\bs{A}^*}^{-1/2}
\left(\bs{I} + \frac{\bs{U}\bs{U}^\top}{1- \|\bs U\|^2}\right ){\bs{A}^*}^{-1/2}\bs{e}_i}{\left({\bs {A}^*}^{-1} \right)_{ii}} - 1  \\ &= \frac{(\bs b_i^\top \bs U)^2}
{1-\|\bs U\|^2} \overset{d}{=} \frac{(\bs{b}_i^\top \bs{Z})^2}{V}.
\end{aligned}
$$

Noting that $\bs{a}_i^\top \bs{b}_i = \sqrt{1- r^2}$, we rotate the spherical $\bs Z$ so that
$\bs a_i^\top\bs Z=Z_1$ and
$\bs b_i^\top\bs Z= \sqrt{1-r^2}Z_1+rZ_2$.
The remaining squared coordinates sum to an independent
$W\sim\chi_{d-2}^2$, proving~\eqref{eq:reference-pair}.
\end{proof}

\subsection{Conditional stochastic ordering and copula concavity}
\label{app:dependence}
\begin{proof}[Proof of Lemma~\ref{lem:copula-concavity}]
Fix $r\in[0,1]$ and put $s=\sqrt{1-r^2}$. Work throughout with the
reference distribution in \eqref{eq:reference-pair}, which corresponds to a fixed
conditional distribution given $\bs A^*$ for the values of $r$ generated
by that matrix. Define
\begin{equation}\label{eq:beta-transforms}
V_1=\frac{Z_1^2}{Z_2^2+W+V},
\qquad
V_2=\frac{(sZ_1+rZ_2)^2}{V}.
\end{equation}
We prove that
$V_2\mid V_1=v_1$ is stochastically nondecreasing in $v_1$ and that
$V_1\mid V_2=v_2$ is stochastically nondecreasing in $v_2$.

\medskip\noindent\emph{Precision-adjusted evidence given marginal evidence.}
Let $R=Z_2^2+W+V$, $M=(Z_2^2+V)/R$, and $Y=Z_2/\sqrt{Z_2^2+V}$.
The factorization of independent chi-squared random variables into sums and proportions shows that $M$, $Y$, and $R$ are mutually independent, with $M\sim\operatorname{Beta}((\nu+1)/2,(d-2)/2)$ for $d > 2$, and $M=1$ with probability $1$ for $d=2$, and $Y$ having density
$f_\nu(y)=\frac{(1-y^2)^{\nu/2-1}}{\mathrm B(1/2,\nu/2)}$ for $-1<y<1$.
Since $Z_1$ is independent of $(M,Y,R)$, the pair $(M,Y)$ is independent of $V_1=Z_1^2/R$.

Writing $\varepsilon=\operatorname{sgn}(Z_1)$, we have $Z_1=\varepsilon\sqrt{V_1R}$, $Z_2=Y\sqrt{MR}$, and $V=(1-Y^2)MR$.
The sign $\varepsilon$ is symmetric and independent of $(V_1,M,Y)$.
Absorbing this sign using the symmetry of $Y$, we obtain $(V_2\mid V_1=v_1)\overset d=\{rY+s\sqrt{v_1/M}\}^{2}/(1-Y^2)$.
Here the joint distribution of $(M,Y)$ does not depend on $v_1$.

For fixed $t>0$, define $F(c;t)=\Pr\{(rY+c)^2/(1-Y^2)\le t\}$.
It suffices to show that $F(c;t)$ is nonincreasing in $c\ge0$, since $c(v_1,m)=s\sqrt{v_1/m}$ is nondecreasing in $v_1$ for every fixed $m>0$, and
$\Pr(V_2\le t\mid V_1=v_1)=E_M\left[F\left(s\sqrt{v_1/M};t\right)\right]$.
Once the stated monotonicity of $F$ is established, averaging over
the unchanged distribution of $M$ shows that this conditional
probability is nonincreasing in $v_1$. Equivalently,
$V_2\mid V_1=v_1$ is stochastically nondecreasing in $v_1$.

The monotonicity of $F$ is immediate if $r=0$. Assume $r >0$, and put $D=r^2+t$ and $q(c)=\sqrt{D-c^2}$.
For $0 \le c < \sqrt{D}$, the inequality inside the probability defining $F$ is equivalent to $DY^2+2rcY+c^2-t\le0$, with the endpoints $\ell(c)=(-rc-\sqrt{t}q(c))/D$ and $h(c)=(-rc+\sqrt{t}q(c))/D$.
Define $w_{\nu}(y)=\sqrt{1-y^2}f_{\nu}(y)=(1-y^2)^{(\nu-1)/2}/\textrm{B}(1/2,\nu/2)$.
Since $\nu\ge1$, $w_\nu$ is symmetric and nonincreasing in $|y|$.
Differentiating the endpoints gives
$$
h'(c)=-\frac{\sqrt{1-h^2(c)}}{q(c)},\qquad
\ell'(c)=
\begin{cases}
-\sqrt{1-\ell^2(c)}/q(c),&0<c<r,\\
\phantom{-}\sqrt{1-\ell^2(c)}/q(c),&r<c<\sqrt D.
\end{cases}
$$
Consequently,
$$
\frac{\partial F(c;t)}{\partial c}
=
\begin{cases}
[w_\nu(\ell(c))-w_\nu(h(c))]/q(c),& 0<c<r,\\[2pt]
[-w_\nu(\ell(c)) -w_\nu(h(c))]/q(c),& r<c<\sqrt D.
\end{cases}
$$
For $0<c<r$, the inequalities
$\ell(c)<h(c)$ and $\ell(c)+h(c)=-2rc/D<0$ imply
$|\ell(c)|>|h(c)|$, so the first expression is nonpositive.
The second expression is nonpositive as well. The density is integrable
and has no atoms, so $F(c;t)$ is continuous at the transition points,
including $c=r$ where $\ell(c)=-1$. Also, $F(c;t)= 0$ for $c \ge \sqrt{D}$. Thus, the desired monotonicity of $F$ holds for all $\nu \ge 1$.


\medskip\noindent\emph{Marginal evidence given precision-adjusted evidence.}


\begin{proposition}
\label{prop:J-anderson-monotonicity}
Let $\nu\ge1$, $r\in[0,1]$, and let $\widetilde Y$ have density
$\widetilde f_\nu(y)=\frac{1}{\mathrm B(1/2,(\nu+1)/2)}(1-y^2)^{(\nu-1)/2}$ for $-1<y<1$.
For each fixed $t>0$, define
$J(c;t)=\Pr\!\left\{\left|c\sqrt{1-\widetilde Y^2}+r\widetilde Y\right|\le t\right\}$ for $c\ge0$.
Then $J(c;t)$ is nonincreasing in $c\ge0$.
\end{proposition}

\begin{proof}
Extend $\widetilde f_\nu$ by zero outside $[-1,1]$. Because $\nu\ge1$,
it is symmetric and nonincreasing in $|y|$, including when $\nu=1$,
for which it is uniform on $(-1,1)$. Thus
Lemma~\ref{lem:anderson-one-dimensional} applies. Write
$g_c(y)=c\sqrt{1-y^2}+ry$ for $-1\le y\le1$.

\medskip\noindent\emph{Case 1: $0<t<r$.}
In this case $r>0$. Set
$D(c)=r^2+c^2$, $m(c)=\frac{c\sqrt{D(c)-t^2}}{D(c)}$, and $h(c)=\frac{rt}{D(c)}$.
Then $\{y\in[-1,1]:|g_c(y)|\le t\}=[-m(c)-h(c),\,-m(c)+h(c)]$.
These quantities are obtained as follows. First, $g_c$ is concave in $y \in [-1,1]$, with
$g_c(-1)=-r<-t$ and $g_c(1)=r>t$. For $c>0$, it first increases
and then decreases, and its decreasing part stays above $t$.
The two boundary crossings therefore lie on the increasing part;
solving $g_c(y)=-t$ and $g_c(y)=t$ there gives respectively
$-m(c)-h(c)$ and $-m(c)+h(c)$. For $c=0$, the formula reduces to
$[-t/r,t/r]$. The interval is contained in $[-1,1]$, since
$m(c)+h(c)=\frac{c\sqrt{D(c)-t^2}+rt}{D(c)}\le1$
by the Cauchy--Schwarz inequality.

As $c$ increases, $c\mapsto h(c)$ is nonincreasing. Moreover,
$m^2(c)=\left(1-\frac{r^2}{D(c)}\right)\left(1-\frac{t^2}{D(c)}\right)$
is nondecreasing, both factors being nonnegative and nondecreasing in
$D(c)$, because $D(c)\ge r^2>t^2$. Thus $m(c)$ is nondecreasing.
For $0\le c_1\le c_2$, symmetry and
Lemma~\ref{lem:anderson-one-dimensional} (in \ref{sec:supp}) now yield
$$
J(c_2;t)= \mathcal M_{\widetilde f_\nu}(m(c_2),h(c_2)) \le \mathcal M_{\widetilde f_\nu}(m(c_1),h(c_2)) \le\mathcal M_{\widetilde f_\nu}(m(c_1),h(c_1)) = J(c_1;t).
$$

\medskip\noindent\emph{Case 2: $t\ge r$.}
For every $c\ge0$ and $y\in[-1,1]$, $g_c(y)\ge-r\ge-t$.
Consequently,
$\{y:|g_c(y)|\le t\}=\{y:g_c(y)\le t\}$.
Since $g_c(y)$ is nondecreasing in $c$ for each fixed $y$,
these acceptance sets are decreasing under set inclusion. Their
probabilities are therefore nonincreasing in $c$. This case also
includes $r=0$.

The two cases together prove the proposition.
\end{proof}

Returning to~\eqref{eq:beta-transforms}, apply the orthogonal transformation
$\widetilde Z_1=rZ_1-sZ_2$, $\widetilde Z_2=sZ_1+rZ_2$.
The rotated variables are independent standard normals, independent of
$(V,W)$. Since $Z_1=r\widetilde Z_1+s\widetilde Z_2$ and the transformation
preserves the sum of squares,
\begin{equation}\label{eq:anderson-reverse-rotated-transforms}
\frac{V_1}{1+V_1}
=
\frac{(r\widetilde Z_1+s\widetilde Z_2)^2}
{\widetilde Z_1^2+\widetilde Z_2^2+W+V},
\qquad
\frac{V_2}{1+V_2}
=
\frac{\widetilde Z_2^2}{\widetilde Z_2^2+V}.
\end{equation}

Put $H=\widetilde Z_2^2+V$, $B=\frac{\widetilde Z_2^2}{H}=\frac{V_2}{1+V_2}$,
$L=\frac{H+\widetilde Z_1^2}{H+\widetilde Z_1^2+W}$, and
$\widetilde Y=\frac{\widetilde Z_1}{\sqrt{H+\widetilde Z_1^2}}$.
The factorization of independent chi-squares into sums and proportions gives
$B\perp(H,\widetilde Z_1,W)$. Applying it again shows that
$L,\widetilde Y,V_2$ are mutually independent, with
$$
L\sim\operatorname{Beta}\left(
\frac{\nu+2}{2},\frac{d-2}{2}
\right)\quad(d>2),
\qquad
\widetilde f_\nu(y)
=
\frac{(1-y^2)^{(\nu-1)/2}}
{\mathrm B(1/2,(\nu+1)/2)},\quad |y|<1,
$$
where $\widetilde f_\nu$ is the density of $\widetilde Y$.
For $d=2$, take $L=1$.

Let $\varepsilon=\operatorname{sgn}(\widetilde Z_2)$, which is symmetric
and independent of $(L,\widetilde Y,V_2)$. Substitution into
\eqref{eq:anderson-reverse-rotated-transforms} yields
$$
\frac{V_1}{1+V_1}
=
L\left\{r\widetilde Y+
s\varepsilon\sqrt B\sqrt{1-\widetilde Y^2}\right\}^{2}.
$$
Conditioning on $V_2=v_2$ fixes $B=v_2/(1+v_2)$ without changing
the distribution of $(L,\widetilde Y,\varepsilon)$. Absorbing $\varepsilon$
by symmetry, using
$(L,\varepsilon\widetilde Y)\overset d=(L,\widetilde Y)$, gives
$$
\frac{V_1}{1+V_1}\,\bigg|\,V_2=v_2
\overset d=
L\left\{c(v_2)\sqrt{1-\widetilde Y^2}
+r\widetilde Y\right\}^{2},
\qquad
c(v_2)=s\sqrt{\frac{v_2}{1+v_2}}.
$$

Since $z\mapsto z/(1+z)$ is increasing and $L>0$ almost surely,
for every $v_1>0$,
$$
\Pr(V_1\le v_1\mid V_2=v_2)
=
\mathbb E_L\left[
J\left(c(v_2);\sqrt{\frac{v_1}{(1+v_1)L}}\right)
\right],
$$
where $J(c;t)=\Pr\left\{\left|c\sqrt{1-\widetilde Y^2}+r\widetilde Y\right|\le t\right\}$.
By Proposition~\ref{prop:J-anderson-monotonicity}, $J(c;t)$ is
nonincreasing in $c$ for every fixed $t>0$. Because $c(v_2)$ is
nondecreasing and the distribution of $L$ does not depend on $v_2$,
averaging preserves this ordering. Hence
$v_2\longmapsto\Pr(V_1\le v_1\mid V_2=v_2)$ is nonincreasing for every $v_1>0$.
Thus $V_1\mid V_2=v_2$ is stochastically nondecreasing in $v_2$.
The argument includes $d=2$ and $r=0$; when $r=1$, $s=0$,
so the conditional law does not depend on $v_2$.

\medskip\noindent\emph{Copula concavity.}
Both $p$-values are decreasing continuous functions of the corresponding $V_j$.
The first ordering makes $p\mapsto\Pr(Q_r^*\le v\mid P_r^*=p)$ nonincreasing.
Since $P_r^*$ is uniform, its integral from zero to $u$, namely $C_r(u,v)$, is concave in $u$.
The reverse ordering gives concavity in $v$.
\end{proof}

\begin{remark}
In the subsection dealing with precision-adjusted evidence given marginal
evidence, the variable $Y$ has the density
$f_\nu(y)=\frac{(1-y^2)^{\nu/2-1}}{\mathrm B(1/2,\nu/2)}$ for $-1<y<1$,
and the required probability is
$F(c;t)=\Pr\!\left\{\frac{(rY+c)^2}{1-Y^2}\le t\right\}$.
For $1\le\nu<2$, this density increases with $|y|$; in particular,
$f_1(y)=1/(\pi\sqrt{1-y^2})$ is U-shaped. It therefore fails the
symmetric-unimodality assumption of
Lemma~\ref{lem:anderson-one-dimensional} that we have used in the reverse-conditioning argument dealing with marginal evidence given precision-adjusted evidence.
\end{remark}

\subsection{Conditional Simes Validity and the Global Test}
\label{app:global-validity}

\begin{proof}[Proof of Theorem~\ref{thm:global-validity}]
For $d\ge2$, fix $\bs {A}^*$ and $i$. Lemma~\ref{lem:conditional-pair} gives uniform conditional marginals, and Lemma~\ref{lem:copula-concavity} gives separate concavity of $C_{r_i(\bs {A}^*)}$. Applying Lemma~\ref{lem:simes-criterion} (in \ref{sec:app-auxiliary-lemmas}) under this conditional law proves~\eqref{eq:pairwise-superuniformity}. For $d=1$, $P_1=Q_1$ is uniform conditional on $A_{11}+X_1^2$. In either case, the conditional union bound gives
$
\Pr(P_{\mathrm{EA}}\le\alpha\mid\bs{A}^*)
\le\sum_{i=1}^d
\Pr\!\left(S_i\le\frac\alpha d\,\middle|\,\bs {A}^*\right)
\le\alpha.
$
Taking expectations over $\bs {A}^*$ proves unconditional validity.
\end{proof}

\bibliographystyle{apalike}
\bibliography{references}

\part*{Supplementary Materials}
\label{sec:supp}
\setcounter{section}{0}\renewcommand{\thesection}{S\arabic{section}}
\setcounter{table}{0}\renewcommand{\thetable}{S\arabic{table}}
\setcounter{figure}{0}\renewcommand{\thefigure}{S\arabic{figure}}
\renewcommand{\theHsection}{S\arabic{section}}\renewcommand{\theHsubsection}{S\arabic{section}.\arabic{subsection}}
\renewcommand{\theHtable}{S\arabic{table}}\renewcommand{\theHfigure}{S\arabic{figure}}

\section{Auxiliary Lemmas}
\label{sec:app-auxiliary-lemmas}

The following lemma gives a bivariate version of Simes' inequality under positive dependence through stochastic ordering (PDS) \citep{Block1985}, also known as positive regression dependence on a subset (PRDS) in the multiple-testing literature \citep[Theorem~1.2]{BenjaminiYekutieli2001}. For uniform marginals, this dependence condition for a bivariate copula is equivalent to separate concavity of the joint distribution function \citep[Corollary~5.2.11]{Nelsen2006}. The lemma thus provides a convenient criterion for the validity of the two-input Simes combination, including under conditional distributions

\begin{lemma}[Pairwise Simes criterion under PDS or PRDS]\label{lem:simes-criterion}
Suppose $P$ and $Q$ have uniform marginals under a probability distribution. Let
$C(u,v)=\Pr(P\le u,Q\le v)$. If $C$ is concave in each argument separately, then $S=\min\{2\min(P,Q),\max(P,Q)\}$ satisfies $\Pr(S\le u)\le u$ for $0\le u\le1$. The statement also applies under a conditional distribution with uniform conditional marginals.
\end{lemma}

The following lemma provides a one-dimensional specialization of the symmetric
convex-set translation inequality of \citet[Theorem~1]{Anderson1955}.

\begin{lemma}\label{lem:anderson-one-dimensional}
Let $f:\mathbb R\to[0,\infty)$ be integrable, symmetric about zero,
and nonincreasing on $[0,\infty)$. For $z\in\mathbb R$ and $h\ge0$, set $\mathcal M_f(z,h)=\int_{z-h}^{z+h}f(y)\,dy$.
Then $\mathcal M_f(z,h)=\mathcal M_f(-z,h)$, and $\mathcal M_f(z,h)$
is nonincreasing in $|z|$ for fixed $h$ and nondecreasing in $h$
for fixed $z$. In particular,
\begin{equation}\label{eq:anderson-moving-shrinking-interval}
|z_2|\ge|z_1|,\qquad 0\le h_2\le h_1
\quad\Longrightarrow\quad
\mathcal M_f(z_2,h_2)\le\mathcal M_f(z_1,h_1).
\end{equation}
\end{lemma}

\section{Null Rejection Intervals for the Matched Pair Design} \label{sec:app-null-rejection-intervals}

The tables from this section use the same null evaluation counts as the power study, together with the independent power study nulls. Rows listing several alternatives refer to the shared null evaluation. 

F1--F5 denote the new correlation groups in increasing correlation order; F6--F9 denote the new dimensions 20, 100, 150 and 200. F5 also supplies the amplitude study and the dimension-50 point. For the EA test in both the settings (Tables \ref{tab:main-null-known} and \ref{tab:main-null-estimated}), the upper bounds cover the level in all cases to showcase non-conservative behavior across designs unlike Z-Max that stays conservatively bound for F7. For M3 in Setting K and C2 in Setting W, all the methods tend to exceed the desired level. Considering the intervals across all designs both in the known and unknown covariance setting, EA intervals contain the nominal level comparably with the other methods.

\begin{table}\centering
\caption{Known-covariance pointwise 95\% Wilson intervals for null rejection probabilities.}\label{tab:main-null-known}
{\footnotesize\setlength{\tabcolsep}{2pt}\renewcommand{\arraystretch}{1.12}
\begin{tabular}{lrrrrrrrrr}
\toprule
Setting & EA & ZM & ZK & IHC & FC & QCQ & ZWS & MP & Q \\
\midrule
M1 & [.043,.052] & [.043,.051] & [.044,.053] & [.047,.055] & [.047,.055] & [.049,.058] & [.048,.057] & [.043,.051] & [.048,.056] \\
M2 & [.042,.050] & [.042,.050] & [.041,.049] & [.046,.055] & [.044,.052] & [.046,.055] & [.046,.054] & [.043,.051] & [.044,.052] \\
M3 & [.051,.060] & [.052,.061] & [.050,.059] & [.045,.054] & [.050,.059] & [.050,.059] & [.050,.059] & [.051,.060] & [.050,.059] \\
M4 & [.046,.055] & [.046,.055] & [.048,.056] & [.045,.054] & [.052,.061] & [.047,.056] & [.052,.061] & [.047,.055] & [.048,.057] \\
M5/M6/M7 & [.045,.054] & [.045,.054] & [.046,.054] & [.045,.053] & [.045,.053] & [.047,.056] & [.045,.053] & [.046,.054] & [.045,.053] \\
M8 & [.047,.056] & [.047,.055] & [.045,.054] & [.042,.050] & [.044,.052] & [.043,.051] & [.042,.050] & [.046,.055] & [.042,.051] \\
M9 & [.048,.057] & [.048,.057] & [.044,.053] & [.044,.052] & [.044,.053] & [.044,.053] & [.044,.053] & [.049,.057] & [.043,.051] \\
B1/B2 & [.045,.053] & [.045,.053] & [.043,.051] & [.042,.050] & [.042,.050] & [.043,.052] & [.042,.050] & [.044,.052] & [.043,.051] \\
B3/B4 & [.042,.051] & [.043,.051] & [.049,.057] & [.049,.057] & [.048,.056] & [.048,.057] & [.047,.055] & [.043,.051] & [.050,.058] \\
C1 & [.049,.058] & [.049,.058] & [.051,.060] & [.047,.055] & [.048,.057] & [.046,.055] & [.046,.055] & [.049,.058] & [.046,.055] \\
C2 & [.048,.056] & [.048,.056] & [.051,.060] & [.047,.055] & [.050,.059] & [.047,.056] & [.052,.061] & [.048,.056] & [.052,.061] \\
C3 & [.045,.054] & [.045,.054] & [.051,.060] & [.048,.056] & [.050,.059] & [.048,.057] & [.048,.057] & [.045,.054] & [.047,.055] \\
F1 & [.049,.058] & [.048,.057] & [.047,.056] & [.041,.050] & [.047,.056] & [.047,.055] & [.048,.056] & [.049,.058] & [.048,.057] \\
F2 & [.043,.051] & [.042,.051] & [.045,.054] & [.049,.058] & [.045,.054] & [.045,.054] & [.044,.053] & [.042,.051] & [.043,.052] \\
F3 & [.047,.056] & [.047,.055] & [.050,.059] & [.049,.058] & [.048,.057] & [.050,.059] & [.048,.057] & [.047,.055] & [.047,.055] \\
F4 & [.046,.054] & [.046,.054] & [.044,.053] & [.048,.057] & [.043,.052] & [.044,.052] & [.043,.051] & [.045,.053] & [.043,.051] \\
F5 & [.044,.052] & [.044,.052] & [.044,.052] & [.046,.054] & [.044,.052] & [.044,.053] & [.042,.050] & [.043,.052] & [.042,.050] \\
F6 & [.048,.057] & [.049,.057] & [.052,.061] & [.045,.053] & [.046,.055] & [.049,.058] & [.044,.053] & [.048,.057] & [.044,.052] \\
F7 & [.042,.050] & [.041,.049] & [.045,.054] & [.045,.054] & [.046,.054] & [.045,.053] & [.048,.057] & [.043,.051] & [.047,.055] \\
F8 & [.047,.056] & [.048,.056] & [.048,.057] & [.049,.058] & [.047,.055] & [.046,.054] & [.049,.058] & [.047,.056] & [.050,.059] \\
F9 & [.044,.052] & [.043,.052] & [.041,.049] & [.047,.056] & [.047,.055] & [.047,.055] & [.046,.055] & [.043,.052] & [.049,.058] \\
\bottomrule
\end{tabular}}
\end{table}
\begin{table}\centering
\caption{Sample covariance pointwise 95\% Wilson intervals for null rejection probabilities .}\label{tab:main-null-estimated}
{\footnotesize\setlength{\tabcolsep}{2pt}\renewcommand{\arraystretch}{1.12}
\begin{tabular}{lrrrrrrrr}
\toprule
Setting & EA & ZM & ZK & FC & QCQ & ZWS & MP & Q \\
\midrule
M1 & [.045,.053] & [.046,.054] & [.046,.055] & [.047,.055] & [.049,.058] & [.048,.057] & [.047,.055] & [.047,.055] \\
M2 & [.042,.051] & [.044,.052] & [.044,.052] & [.044,.052] & [.046,.055] & [.046,.054] & [.045,.053] & [.040,.048] \\
M3 & [.049,.058] & [.050,.059] & [.053,.062] & [.050,.059] & [.050,.059] & [.050,.059] & [.050,.059] & [.048,.057] \\
M4 & [.044,.052] & [.047,.056] & [.048,.056] & [.052,.061] & [.047,.056] & [.052,.061] & [.047,.055] & [.051,.060] \\
M5/M6/M7 & [.048,.057] & [.046,.055] & [.045,.054] & [.045,.053] & [.047,.056] & [.045,.053] & [.047,.056] & [.045,.054] \\
M8 & [.044,.053] & [.047,.055] & [.046,.054] & [.044,.052] & [.043,.051] & [.042,.050] & [.046,.054] & [.041,.049] \\
M9 & [.049,.058] & [.047,.055] & [.045,.054] & [.044,.053] & [.044,.053] & [.044,.053] & [.047,.056] & [.043,.051] \\
B1/B2 & [.043,.051] & [.040,.048] & [.041,.049] & [.042,.050] & [.043,.052] & [.042,.050] & [.041,.049] & [.044,.052] \\
B3/B4 & [.043,.052] & [.044,.052] & [.049,.058] & [.048,.056] & [.048,.057] & [.047,.055] & [.044,.053] & [.047,.055] \\
C1 & [.047,.056] & [.048,.057] & [.049,.058] & [.048,.057] & [.046,.055] & [.046,.055] & [.049,.058] & [.043,.051] \\
C2 & [.053,.062] & [.047,.055] & [.052,.061] & [.050,.059] & [.047,.056] & [.052,.061] & [.047,.055] & [.053,.063] \\
C3 & [.048,.056] & [.048,.057] & [.051,.060] & [.050,.059] & [.048,.057] & [.048,.057] & [.047,.056] & [.046,.054] \\
F1 & [.049,.058] & [.047,.056] & [.046,.055] & [.047,.056] & [.047,.055] & [.048,.056] & [.047,.056] & [.045,.054] \\
F2 & [.045,.053] & [.043,.051] & [.047,.055] & [.045,.054] & [.045,.054] & [.044,.053] & [.043,.052] & [.047,.056] \\
F3 & [.049,.058] & [.047,.055] & [.050,.059] & [.048,.057] & [.050,.059] & [.048,.057] & [.047,.055] & [.046,.055] \\
F4 & [.044,.053] & [.044,.053] & [.044,.053] & [.043,.052] & [.044,.052] & [.043,.051] & [.045,.053] & [.044,.052] \\
F5 & [.042,.050] & [.045,.054] & [.043,.051] & [.044,.052] & [.044,.053] & [.042,.050] & [.045,.053] & [.042,.050] \\
F6 & [.044,.053] & [.047,.055] & [.048,.057] & [.046,.055] & [.049,.058] & [.044,.053] & [.046,.055] & [.044,.052] \\
F7 & [.042,.050] & [.042,.050] & [.045,.053] & [.046,.054] & [.045,.053] & [.048,.057] & [.043,.051] & [.046,.054] \\
F8 & [.047,.055] & [.046,.055] & [.048,.057] & [.047,.055] & [.046,.054] & [.049,.058] & [.046,.055] & [.046,.055] \\
F9 & [.044,.052] & [.043,.051] & [.043,.051] & [.047,.055] & [.047,.055] & [.046,.055] & [.043,.051] & [.048,.057] \\
\bottomrule
\end{tabular}}
\end{table}

\section{A Broader Simulation Design}\label{sec:app-broad-study}

We 
report a broader study, varying covariance strength, dimension and amplitude, following the protocol from Sections~\ref{sec:methods} and \ref{sec:mc}. 
All panels use \(B=19,999\), \(R=10,000\) and level $\alpha=.05$. 

\subsection{Correlation, Dimension and Amplitude Settings}
For the correlation design, $d=50$, $n=249$,
$\bs\Sigma=\operatorname{diag}\{\bs R_5(\rho),\bs R_{45}(.1)\}$,
$\bs\mu=2\bs e_1$, and $\rho\in\{.1,.3,.5,.7,.9\}$.
For the amplitude design, retain this covariance with \(\rho=.9\) and vary \(\bs \mu=a \bs e_1\), \(a \in \{1,1.5,2,2.5,3\}\). For the dimension design, take $d\in\{50,100,150,200,250\}$ and $n=3d-1$, with
$$
\bs\Sigma=\operatorname{diag}\{\bs R_{10}(.7),\underbrace{\bs R_{10}(.1),\ldots,\bs R_{10}(.1)}_{(d/10)-1}\},
$$
and $\bs\mu=(3,\underbrace{.5,\ldots,.5}_{9},0,\ldots,0)^\top$.
Thus the dimension design changes both \(d\) and \(n\), while preserving the active ten-coordinate block.  Table~\ref{tab:supp-settings} identifies the 14 distinct alternatives displayed in the power tables.

\begin{table}\centering
\caption{Fixed sparse-heavy locally correlated settings.}\label{tab:supp-settings}
{\small\setlength{\tabcolsep}{8pt}\renewcommand{\arraystretch}{1.12}
\begin{tabular}{llrrrr}
\toprule
Setting & Variable & $d$ & $n$ & Active $\rho$ & $a$ \\
\midrule
O1 & Correlation & 50 & 249 & 0.1 & 2 \\
O2 & Correlation & 50 & 249 & 0.3 & 2 \\
O3 & Correlation & 50 & 249 & 0.5 & 2 \\
O4 & Correlation & 50 & 249 & 0.7 & 2 \\
O5 & Correlation & 50 & 249 & 0.9 & 2 \\
O6 & Amplitude & 50 & 249 & 0.9 & 1 \\
O7 & Amplitude & 50 & 249 & 0.9 & 1.5 \\
O8 & Amplitude & 50 & 249 & 0.9 & 2.5 \\
O9 & Amplitude & 50 & 249 & 0.9 & 3 \\
O10 & Dimension & 50 & 149 & 0.7 & 3 \\
O11 & Dimension & 100 & 299 & 0.7 & 3 \\
O12 & Dimension & 150 & 449 & 0.7 & 3 \\
O13 & Dimension & 200 & 599 & 0.7 & 3 \\
O14 & Dimension & 250 & 749 & 0.7 & 3 \\
\bottomrule
\end{tabular}}
\end{table}

\subsection{Power Comparisons for Broader Designs} \label{app:broad-design}
\begin{table}\centering
\caption{Power comparisons in a broader design study for known covariance. Bold denotes the largest displayed point estimate.}\label{tab:supp-known}
{\small\setlength{\tabcolsep}{3pt}\renewcommand{\arraystretch}{1.12}
\begin{tabular}{lrrrrrrrrr}
\toprule
Setting & EA & ZM & ZK & IHC & FC & QCQ & ZWS & MP & Q \\
\midrule
O1 & 0.151 & 0.155 & \textbf{0.158} & 0.0890 & 0.119 & 0.0923 & 0.116 & 0.146 & 0.113 \\
O2 & 0.174 & 0.183 & \textbf{0.185} & 0.0953 & 0.116 & 0.0905 & 0.110 & 0.145 & 0.125 \\
O3 & 0.262 & \textbf{0.287} & 0.283 & 0.113 & 0.107 & 0.0868 & 0.103 & 0.141 & 0.161 \\
O4 & 0.504 & \textbf{0.545} & 0.526 & 0.152 & 0.109 & 0.0849 & 0.106 & 0.151 & 0.282 \\
O5 & 0.989 & \textbf{0.993} & 0.989 & 0.280 & 0.103 & 0.0835 & 0.0972 & 0.148 & 0.832 \\
O6 & 0.333 & \textbf{0.376} & 0.362 & 0.126 & 0.0523 & 0.0538 & 0.0510 & 0.0582 & 0.193 \\
O7 & 0.812 & \textbf{0.844} & 0.818 & 0.195 & 0.0656 & 0.0641 & 0.0618 & 0.0829 & 0.499 \\
O8 & \textbf{1.00} & \textbf{1.00} & \textbf{1.00} & 0.395 & 0.179 & 0.112 & 0.166 & 0.263 & 0.977 \\
O9 & \textbf{1.00} & \textbf{1.00} & \textbf{1.00} & 0.521 & 0.308 & 0.156 & 0.280 & 0.428 & 0.999 \\
O10 & 0.851 & \textbf{0.870} & 0.835 & 0.161 & 0.291 & 0.180 & 0.278 & 0.421 & 0.524 \\
O11 & 0.803 & \textbf{0.824} & 0.751 & 0.136 & 0.273 & 0.187 & 0.266 & 0.360 & 0.368 \\
O12 & 0.777 & \textbf{0.791} & 0.680 & 0.128 & 0.257 & 0.217 & 0.256 & 0.317 & 0.293 \\
O13 & 0.753 & \textbf{0.780} & 0.624 & 0.115 & 0.235 & 0.218 & 0.238 & 0.292 & 0.250 \\
O14 & 0.736 & \textbf{0.761} & 0.569 & 0.120 & 0.220 & 0.206 & 0.225 & 0.277 & 0.212 \\
\bottomrule
\end{tabular}}
\end{table}
\begin{table}\centering
\caption{Power comparisons in a broader design study for known covariance. Bold denotes the largest displayed point estimate.}\label{tab:supp-estimated}
{\small\setlength{\tabcolsep}{3pt}\renewcommand{\arraystretch}{1.12}
\begin{tabular}{lrrrrrrrr}
\toprule
Setting & EA & ZM & ZK & FC & QCQ & ZWS & MP & Q \\
\midrule
O1 & 0.141 & 0.145 & 0.136 & 0.119 & 0.0923 & 0.116 & \textbf{0.146} & 0.103 \\
O2 & \textbf{0.151} & 0.149 & 0.144 & 0.116 & 0.0905 & 0.110 & 0.136 & 0.110 \\
O3 & \textbf{0.209} & 0.198 & 0.185 & 0.107 & 0.0868 & 0.103 & 0.139 & 0.146 \\
O4 & \textbf{0.360} & 0.315 & 0.289 & 0.109 & 0.0849 & 0.106 & 0.147 & 0.240 \\
O5 & \textbf{0.948} & 0.712 & 0.652 & 0.103 & 0.0835 & 0.0972 & 0.143 & 0.743 \\
O6 & \textbf{0.231} & 0.0703 & 0.0788 & 0.0523 & 0.0538 & 0.0510 & 0.0554 & 0.160 \\
O7 & \textbf{0.652} & 0.273 & 0.253 & 0.0656 & 0.0641 & 0.0618 & 0.0805 & 0.403 \\
O8 & \textbf{0.998} & 0.961 & 0.937 & 0.179 & 0.112 & 0.166 & 0.252 & 0.948 \\
O9 & \textbf{1.00} & 0.999 & 0.997 & 0.308 & 0.156 & 0.280 & 0.412 & 0.996 \\
O10 & 0.614 & \textbf{0.682} & 0.626 & 0.291 & 0.180 & 0.278 & 0.407 & 0.381 \\
O11 & 0.562 & \textbf{0.638} & 0.554 & 0.273 & 0.187 & 0.266 & 0.349 & 0.276 \\
O12 & 0.530 & \textbf{0.626} & 0.489 & 0.257 & 0.217 & 0.256 & 0.312 & 0.226 \\
O13 & 0.492 & \textbf{0.617} & 0.440 & 0.235 & 0.218 & 0.238 & 0.290 & 0.185 \\
O14 & 0.467 & \textbf{0.594} & 0.394 & 0.220 & 0.206 & 0.225 & 0.272 & 0.161 \\
\bottomrule
\end{tabular}}
\end{table}

The broader study gives a different picture from the examples in Section \ref{subsec:mainsettings}. In the varying correlation design, only the first coordinate has a signal; its correlated neighbors have zero means. As correlation increases, those neighbors become increasingly useful for removing shared noise from the first coordinate. Precision adjustment therefore amplifies the standardized signal strongly. In the matched-pair designs, the mean vector itself changes with correlation and together with the precision-adjusted mean provide complementary evidence.

\begin{figure}
\centering
\begin{minipage}{0.32\textwidth}
\centering

\includegraphics[width=\linewidth]
{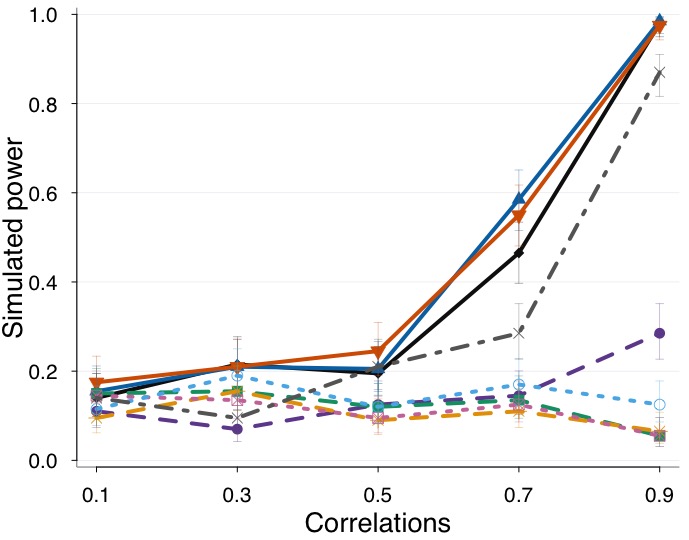}
\end{minipage}
\hfill
\begin{minipage}{0.32\textwidth}
\centering

\includegraphics[width=\linewidth]
{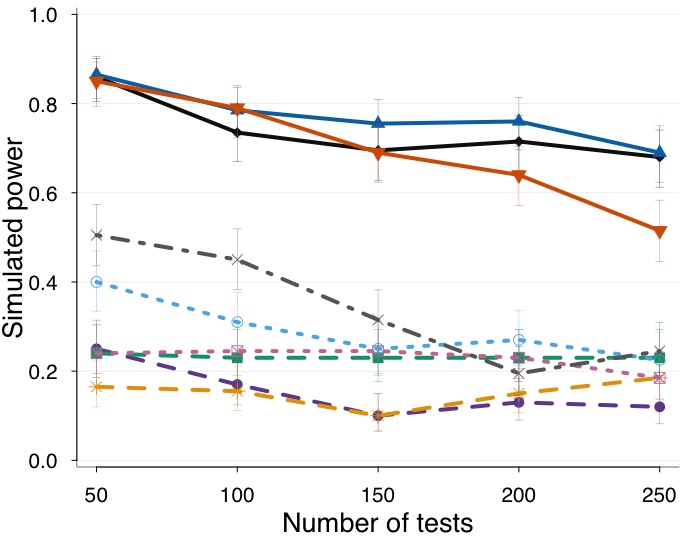}
\end{minipage}
\hfill
\begin{minipage}{0.32\textwidth}
\centering

\includegraphics[width=\linewidth]
{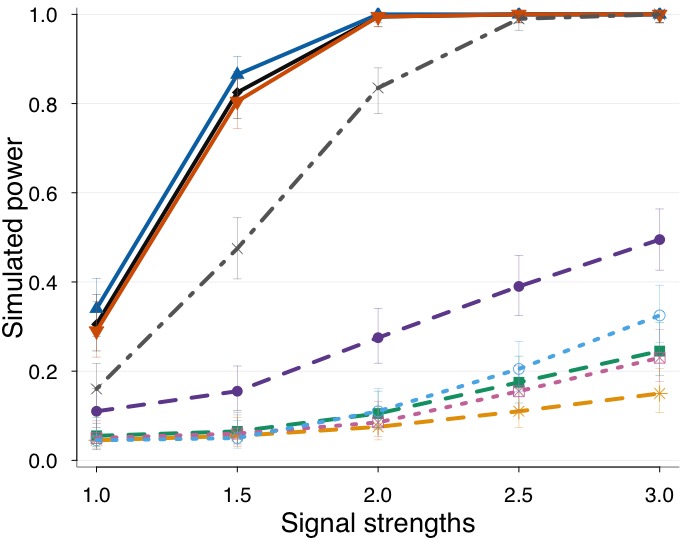}
\end{minipage}

\vspace{1em}

\begin{minipage}{0.32\textwidth}
\centering

\includegraphics[width=\linewidth]
{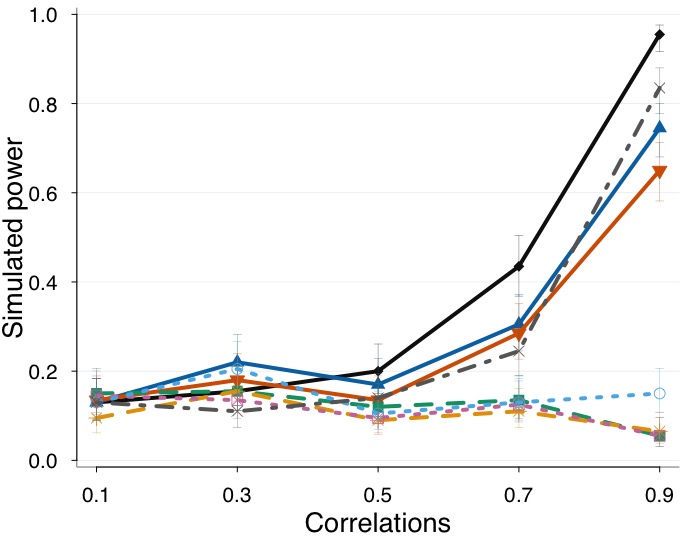}
\end{minipage}
\hfill
\begin{minipage}{0.32\textwidth}
\centering
\includegraphics[width=\linewidth]
{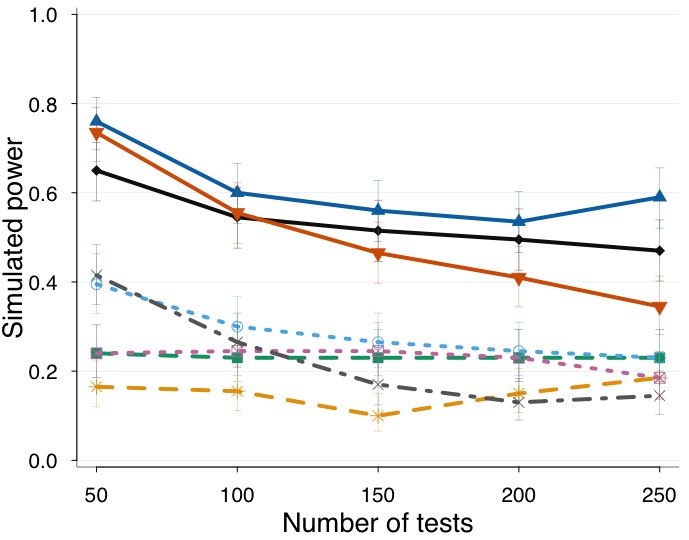}
\end{minipage}
\hfill
\begin{minipage}{0.32\textwidth}
\centering
\includegraphics[width=\linewidth]
{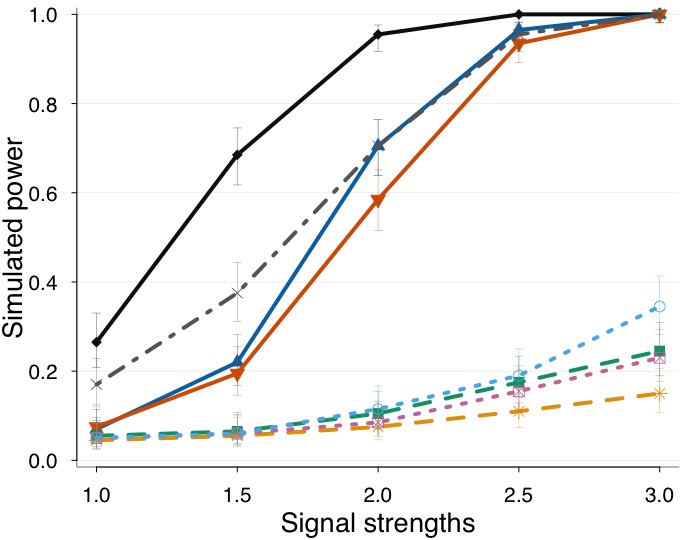}
\end{minipage}

\caption{Simulated power in known-covariance (top row) and sample covariance (bottom row) settings in the sparse, locally correlated designs. From left to right, the horizontal axes vary in correlation, dimension and amplitude. Whiskers are pointwise 95\% Wilson intervals; $B=1999$, $R=2000$, $\alpha=.05$. Methods compared are EA (filled diamond, black); Z-Max (filled upward triangle, blue); Z-TopK (filled downward triangle, vermilion); IHC (filled circle, purple); Feng-COM (filled square, green); QCQ (asterisk, orange); ZWS (crossed square, reddish purple); MinP (open circle, sky blue); Quadratic (cross, gray). }
\label{fig:supp-text}
\end{figure}

Figure~\ref{fig:supp-text} shows that the Z-Max and Z-TopK methods can give better power in a variety of scenarios, while the  quadratic test also performs well. One possible explanation is that this setting has an isolated signal strength in the precision-adjusted mean and Z-Max acts more sensitively to it. For instance, the dimension design has active marginal means {\((3,.5,\ldots,.5)\)} but after precision-adjustment, the vector contains {\((4.379, -.421,\ldots,-.421)\)}. This setting favors Z-Max.

The difference is whether marginal and precision-adjusted statistics provide complementary representations, 
or whether one is already enough. 
For instance, at {\(\rho=.9,\ a=2\)}, the population effects are marginally {$m=(2,0,0,0,0)$} but the precision-adjusted effects are $pa=(5.672, -1.380$, $-1.380, -1.380,-1.380)$; so in this case the information is somewhat redundant. The EA test requires additional structure to have a prominent effect in both the evidence arms. For instance, in the matched design, adjustment preserves the absolute standardized effects, and the EA test can leverage this efficiently.

\section{Additional Simulation Checks}\label{sec:app-new-settings-check}
The settings in Sections~\ref{subsec:correlated-module} and~\ref{subsec:moving average} use separate random draws for exploration, setting critical values, null evaluation, and power evaluation. The methods share the same observations within each replicate. The three amplitudes share the underlying noise and covariance draws. All covariance estimates and nodewise Lasso fits are recomputed for every new replicate.

To generate the observation-level inputs for QCQ and ZWS efficiently, we draw $N=n+1$ independent zero-mean Gaussian rows, center them to obtain $\bs C$, and set $\bs A=\bs C^\top\bs C$. Independently, we generate $\bs X\sim N_d(\bs\mu,\bs\Sigma)$ and form rows $\bs R_i=\bs C_i+\bs X/\sqrt{N}$. This has the same joint distribution as the independent random-orientation construction in Section~\ref{subsec:simulation-methods}. The observed null rejection rates range from $0.0432$ to $0.0539$. This is partly due to the finite number of draws used to set the Monte Carlo critical value.

\begin{table}\centering
\caption{Independent null rejection rates for the supporting settings beyond matched pair. The nominal level is $.05$.}\label{tab:new-settings-null}
{\small\setlength{\tabcolsep}{4pt}\renewcommand{\arraystretch}{1.12}

\begin{tabular}{lrrrrrrrrr}
\toprule
Setting & EA & ZM & ZK & IHC & FC & QCQ & ZWS & MP & Q \\
\midrule
\multicolumn{10}{l}{\emph{Known covariance}}\\
Correlated module & 0.0478 & 0.0472 & 0.0465 & 0.0530 & 0.0477 & 0.0487 & 0.0486 & 0.0480 & 0.0486 \\
Moving average peak & 0.0471 & 0.0482 & 0.0503 & 0.0432 & 0.0539 & 0.0532 & 0.0527 & 0.0492 & 0.0494 \\
\midrule
\multicolumn{10}{l}{\emph{Estimated covariance}}\\
Correlated module & 0.0483 & 0.0485 & 0.0489 & --- & 0.0477 & 0.0487 & 0.0486 & 0.0489 & 0.0506 \\
Moving average peak & 0.0513 & 0.0485 & 0.0469 & --- & 0.0539 & 0.0532 & 0.0527 & 0.0506 & 0.0488 \\
\bottomrule
\end{tabular}}
\end{table}

\paragraph{Uncertainty in power differences.}
We estimate Monte Carlo uncertainty by a two-stage paired bootstrap with $2000$ resamples. Calibration rows are resampled jointly across methods and covariance panels, and independent power evaluation rows are resampled jointly across methods and amplitudes. In every resample we recompute the null-rank thresholds. This includes both uncertainty in setting critical values and the paired sampling variation of the power differences. For each difference $\widehat\Delta$ with bootstrap standard error $s$, we compute the one-sided lower bound $\widehat\Delta-z_{1-.05/90}s$, where $z_{1-.05/90}\approx3.261$. The Bonferroni factor covers all $90$ comparisons: two settings, three amplitudes, eight known-covariance comparators and seven estimated-covariance comparators. All $90$ simultaneous lower bounds are positive; the smallest is $0.0188$.

\paragraph{Analytic EA cutoff.}
For completeness, Table~\ref{tab:new-settings-analytic} evaluates the finite-sample rule $P_{\mathrm{EA}}\le.05$ without Monte Carlo recalibration of EA. Its observed null rejection rates are $.0460$ and $.0392$ for the correlated module in Settings K and W, respectively, and $.0407$ and $.0426$ for the moving average peak. The point estimates of its power also exceed those of every Monte Carlo calibrated comparator in these rows.
\begin{table}\centering
\caption{Power of analytically calibrated EA and the strongest Monte Carlo calibrated comparator. Each entry uses $10,000$ power replicates.}\label{tab:new-settings-analytic}
{\small\setlength{\tabcolsep}{6pt}\renewcommand{\arraystretch}{1.12}

\begin{tabular}{lrrrrr}
\toprule
Setting & $a$ & \multicolumn{2}{c}{Known covariance} & \multicolumn{2}{c}{Estimated covariance} \\
\cmidrule(lr){3-4}\cmidrule(lr){5-6}
 & & Analytic EA & Best baseline & Analytic EA & Best baseline \\
\midrule
Correlated module & 3 & 0.467 & 0.421 & 0.432 & 0.420 \\
Correlated module & 3.5 & 0.665 & 0.603 & 0.630 & 0.601 \\
Correlated module & 4 & 0.831 & 0.771 & 0.799 & 0.767 \\
\midrule
Moving average peak & 3 & 0.494 & 0.456 & 0.463 & 0.446 \\
Moving average peak & 3.5 & 0.698 & 0.637 & 0.661 & 0.646 \\
Moving average peak & 4 & 0.860 & 0.799 & 0.833 & 0.810 \\
\bottomrule
\end{tabular}}
\end{table}

\end{document}